\magnification=1200
\hoffset=.0cm
\voffset=.0cm
\baselineskip=.55cm plus .55mm minus .55mm

%
%
%
\input amssym.def\input amssym.tex
%
%
%
%
%
%
%
%
%


\font\grassettogreco=cmmib10
\font\scriptgrassettogreco=cmmib7
\font\scriptscriptgrassettogreco=cmmib10 at 5 truept
\textfont13=\grassettogreco
\scriptfont13=\scriptgrassettogreco
\scriptscriptfont13=\scriptscriptgrassettogreco


\font\sansserif=cmss10
\font\scriptsansserif=cmss10 at 7 truept
\font\scriptscriptsansserif=cmss10 at 5 truept
\textfont14=\sansserif
\scriptfont14=\scriptsansserif
\scriptscriptfont14=\scriptscriptsansserif


\font\capital=rsfs10
\font\scriptcapital=rsfs10 at 7 truept
\font\scriptscriptcapital=rsfs10 at 5 truept
\textfont15=\capital
\scriptfont15=\scriptcapital
\scriptscriptfont15=\scriptscriptcapital
\def\scri{\fam=15}


\font\euler=eusm10
\font\scripteuler=eusm7
\font\scriptscripteuler=eusm5 
\textfont12=\euler
\scriptfont12=\scripteuler
\scriptscriptfont12=\scriptscripteuler
\def\eul{\fam=12}

%

%

%
%
%
%
%

%
%
%
%
%
%
\def\ref#1{\lbrack#1\rbrack}
%
%
%
%
%
\def\dim{{\rm dim}\hskip 1pt}

\def\deg{{\rm deg}\hskip 1pt}

\def\SU{{\rm SU}}

\def\End{{\rm End}\hskip 1pt}
\def\Fun{{\rm Fun}\hskip 1pt}

\def\hst1{\hskip 1pt}

%
%
%
%
%

\def\titlebf#1{\vskip.5cm${\underline{\hbox{\bf #1}}}$\vskip.5cm}

\def\xxx{\phantom{xxxxxxxxxxxxxxxxxxxxxxxxxxxxx}}


\hrule\vskip.5cm
\hbox to 16.5 truecm{January 2005 \hfil DFUB 05--01}
\hbox to 16.5 truecm{Version 2  \hfil hep-th/0501062}
\vskip.5cm\hrule
\vskip.9cm
\centerline{\bf GENERALIZED COMPLEX GEOMETRY, GENERALIZED BRANES}   
\centerline{\bf AND THE HITCHIN SIGMA MODEL}   
\vskip.4cm
\centerline{by}
\vskip.4cm
\centerline{\bf Roberto Zucchini}
\centerline{\it Dipartimento di Fisica, Universit\`a degli Studi di Bologna}
\centerline{\it V. Irnerio 46, I-40126 Bologna, Italy}
\centerline{\it I.N.F.N., sezione di Bologna, Italy}
\centerline{\it E--mail: zucchinir@bo.infn.it}
\vskip.9cm
\hrule
\vskip.6cm
\centerline{\bf Abstract} 
\vskip.4cm
\par\noindent
Hitchin's generalized complex geometry has been shown to be relevant in 
compactifications of superstring theory with fluxes and is expected to lead to 
a deeper understanding of mirror symmetry. Gualtieri's notion 
of generalized complex submanifold seems to be a natural candidate
for the description of branes in this context. 
Recently, we introduced a Batalin--Vilkovisky field theoretic 
realization of generalized complex geometry, the Hitchin sigma model,
extending the well known Poisson sigma model. In this paper, exploiting Gualtieri's 
formalism, we incorporate branes into the model. A detailed study of the boundary 
conditions obeyed by the world sheet fields is provided. Finally, it is found that, 
when branes are present, the classical Batalin--Vilkovisky cohomology
contains an extra sector that is related non trivially to a novel
cohomology associated with the branes as generalized complex submanifolds. 
\vskip.6cm
\hrule
\vskip.6cm
\par\noindent
MSC-class: 53D17, 53B50. Keywords: Poisson Sigma Model, Generalized Complex Geometry,
Cohomology.
\vfill\eject

\titlebf{Contents}   

\item{1.} Introduction

\item{2.} Generalized complex geometry

\item{3.} Generalized complex submanifolds and branes

\item{4.} 2--dimensional de Rham superfields

\item{5.} The Hitchin sigma model in the presence of branes

\item{6.} The classical BV cohomology in the presence of branes

\item{7.} Discussion

\vfill\eject

\titlebf{1. Introduction}

\par
Mirror symmetry is a duality relating compactifications of type IIA and type IIB 
superstring theory, which yield the same four--dimensional low energy 
effective theory. It has played an important role in the study of 
Calabi--Yau compactifications for both its theoretical implications 
and practical usefulness. Recently, more general compactifications allowing 
for non Ricci--flat metrics and NSNS and RR fluxes have been object of intense 
inquiry. The natural question arises about whether mirror symmetry 
generalizes to this broader class of compactifications and, if so, which its 
properties are. 
This program was outlined originally in refs. \ref{1,2} and was subsequently 
implemented with an increasing degree of generality in a series of papers
\ref{3--5}. These studies indicate that mirror symmetry can be defined for a 
class of manifolds with $\SU(3)$ structure. In spite of these advancements, 
certain aspects of mirror symmetry remain mysterious.

In 2002, Hitchin formulated the notion of generalized complex geometry, which
at the same time unifies and extends the customary notions of complex and 
symplectic geometry and incorporates a natural generalization of Calabi--Yau 
geometry \ref{6}. Hitchin's ideas were developed by Gualtieri \ref{7},
who also worked out the theory of generalized Kaehler geometry. 
The $\SU(3)$ structure manifolds considered in \ref{4,5} are generalized 
Calabi--Yau manifolds as defined by Hitchin. This indicates that 
generalized complex geometry may provide the mathematical set up appropriate  
for the study of mirror symmetry for general flux compactifications.

Type II superstring Calabi--Yau compactifications are described by $(2,2)$ 
superconformal sigma models with Calabi--Yau target manifolds. These field theories
are however rather complicated and, so, they are difficult to study. 
In 1988, Witten showed that a $(2,2)$ supersymmetric sigma model on a 
Calabi--Yau space could be twisted in two different ways,   
to give the so called A and B topological sigma models \ref{8}.  
Unlike the original untwisted sigma model, the topological models are soluble:
the calculation of observables can be reduced to classical problems of geometry.
For this reason, the topological models constitute an ideal field theoretic ground
for the study of mirror symmetry. 

Mirror symmetry relates the A and B models with mirror target manifolds \ref{9}.
The A and B models depend only on the symplectic and complex geometry of the target 
manifold, respectively. Therefore, generalized complex geometry, which 
unifies these two types of geometry, may provide a natural mathematical framework for 
a unified understanding of them. It is also conceivable that topological sigma models 
with generalized Kaehler targets may exhibit the form of generalized mirror symmetry 
encountered in flux compactifications of superstring theory \ref{10--15}. 

D--branes are extended solitonic objects of the superstring spectrum, which are 
expected to play an important role in the ultimate non perturbative understanding 
of superstring physics. They appear as hypersurfaces in space--time where the 
ends of open strings are constrained to lie. 
D--branes appear in a $(2,2)$ superconformal sigma model as certain boundary 
conditions for the world sheet fields. Topological branes appear 
in the associated topological sigma models again as boundary conditions. 
The branes of the A and B model are called A-- and B--branes, respectively. 
Expectedly, mirror symmetry exchanges the sets of A--branes and B--branes
of mirror manifolds. This constitutes a major motivation for their study.

In 1994, Kontsevich formulated the homological mirror symmetry conjecture \ref{16},
which interprets mirror symmetry as the equivalence of two triangulated categories: 
the bounded derived category of coherent sheaves and the derived 
Fukaya category of graded Lagrangian submanifolds carrying flat vector bundles. 
After the emergence of D--branes in superstring theory, 
it was argued that the topological B--branes formed a category which  
could be identified with the derived category of coherent sheaves  \ref{17,18}. 
Since mirror symmetry exchanges A--branes and B--branes, one would expect that the
topological A--branes also form a category and that this category could be 
identified with the derived Fukaya category. This belief was supported by 
Witten's original work \ref{19}, where A--branes appeared as 
Lagrangian submanifolds, and by the analysis of ref. \ref{20}, which showed that 
ghost number anomaly cancellation required these Lagrangian submanifolds to be 
graded. However, other studies indicated that 
there could be A--branes which were not Lagrangian submanifolds \ref{21}. 
The careful analysis of \ref{22} showed that 
a class of coisotropic submanifolds carrying non trivial line bundles
could also serve as A--branes, at least at the classical level. 
This finding suggests that the category of A--branes should be an appropriate 
extension of the derived Fukaya category including the coisotropic branes. 

For reasons explained above, it is expected that generalized complex geometry 
may furnish the appropriate framework for the understanding of mirror symmetry 
also in the presence of branes. In this case, the notion of generalized 
complex submanifold formulated by Gualtieri in \ref{7} should play an important 
role. In fact, it includes as particular cases all known examples of topological 
branes, including the coisotropic ones. However, to the best of our knowledge, 
not much has been tried in this direction so far \ref{10,15,23}.

Any attempt to understand mirror symmetry in the light of generalized complex 
geometry unavoidably must go through some sigma model realization of this latter.
Several such realizations have been proposed so far \ref{10,13,23--32}. 
They have remarkable properties and are also interesting in their own. 
This paper is a further step along this line of development, as we shall 
describe next. 

To a varying degree, the sigma model realizations of generalized complex 
geometry are all related to the well--known Poisson sigma model \ref{33,34}.
This is not surprising, in view of the fact that a generalized complex manifold is 
also a Poisson manifold. 
In \ref {31}, adapting and generalizing the formulation of the 
Poisson sigma model of ref. \ref{35}, based on the Batalin--Vilkovisky quantization 
algorithm \ref{36,37}, we formulated a new realization called Hitchin sigma model. 
We showed that the algebraic properties and integrability conditions of the target space 
generalized almost complex structure are sufficient, non necessary conditions 
for the fulfillment of the Batalin--Vilkovisky classical master equation $(S,S)=0$. 
Further, a non trivial relation between a sector of the classical Batalin--Vilkovisky 
cohomology, on one hand, and a generalized Dolbeault cohomology and the cohomology of 
the generalized deformation complex, on the other, was found. 

In this paper, we continue the study of Hitchin sigma model on a generalized 
complex manifold and show how generalized branes can be incorporated into it. 
We argue that branes are aptly described by Gualtieri's theory of 
generalized complex submanifolds \ref{7}. In the presence of a brane, the 
world sheet has a non empty boundary. The fields therefore must obey appropriate 
boundary conditions, which are determined and analyzed in detail in the paper.
We show that the algebraic properties and integrability 
conditions of the target space generalized almost complex structure as well as 
the algebraic properties of the brane as generalized complex submanifold 
are sufficient, non necessary conditions for the fulfillment of 
the Batalin--Vilkovisky classical master equation $(S,S)=0$. 
The compatibility of the boundary conditions with the so called $b$ symmetry 
and with the overall classical Batalin--Vilkovisky cohomological structure is 
ascertained. 
Further, we find that, when branes are present, the Batalin--Vilkovisky cohomology
contains an extra sector that is related non trivially to an hitherto unknown 
cohomology associated with the branes as generalized complex submanifolds. 
Finally, we compare our approach with the Alexandrov--Kontsevich--Schwartz--Zaboronsky 
formulation of the Poisson sigma model without and with branes worked out in
refs. \ref{38,39}.

The plan of this paper is as follows. In sect. 2, we review the basic notions
of twisted generalized complex geometry. In sect. 3, we illustrate the 
theory of generalized complex submanifolds and show its use for the 
description of generalized branes. In sect. 4, we introduce 
the 2--dimensional de Rham superfield formalism for world sheets with boundary.
In sect. 5, we review the twisted Hitchin sigma model in the absence of branes 
and then show how branes can be incorporated by imposing suitable 
boundary conditions on the fields. In sect. 6, we analyze
the compatibility of the boundary conditions 
with the Batalin--Vilkovisky cohomological structure when branes are present
and describe in detail the relation between the Batalin--Vilkovisky cohomology
and the generalized complex submanifold cohomology of the branes. 
Finally, in sect. 7, we compare our results with related works on the 
Poisson sigma model with coisotropic branes.  


\titlebf{2. Generalized complex geometry}

\par
The notion of generalized complex structure was introduced by Hitchin 
in \ref{6} and developed by Gualtieri \ref{7} in his thesis. 
It encompasses the usual notions of complex and symplectic 
structure as special cases. 
It is the complex counterpart of the notion of Dirac structure, introduced by 
Courant and Weinstein, which unifies Poisson and symplectic geometry \ref{40,41}. 
In this section, we review the basic definitions and results of generalized complex 
geometry used in the sequel of the paper.

Let $M$ be a manifold. In what follows, $M$ will always be 
of even dimension $d$, since the basic structures of generalized complex 
geometry exist only in this case. 

The constructions of generalized complex geometry are all based on 
the vector bundle $TM\oplus T^*M$. 
A generic section $X+\xi\in C^\infty(TM\oplus T^*M)$ 
of this bundle is the direct sum of a vector field $X\in C^\infty(TM)$
and a $1$--form $\xi\in C^\infty(T^*M)$.

$TM\oplus T^* M$ is equipped with a natural indefinite metric of signature  
$(d,d)$ defined by 
$$
\langle X+\xi,Y+\eta\rangle=\hbox{$1\over 2$}(i_X\eta+i_Y\xi),
\eqno(2.1)
$$
for $X+\xi, Y+\eta\in C^\infty(TM\oplus T^* M)$, where $i_V$ 
denotes contraction with respect to a vector field $V$. 
This metric has a large isometry group. This contains the full 
diffeomorphism group of $M$, acting by pull--back. It also contains
the following distinguished isometries, called $b$ transforms, 
defined by 
$$
\exp(b)(X+\xi)= X+\xi+i_X b,
\eqno(2.2)
$$
where $b\in C^\infty(\wedge^2 T^* M)$ is a $2$--form. As it turns out,
$b$ transformation is the most basic symmetry of generalized complex geometry.

An $H$ field is a closed $3$--form $H\in C^\infty(\wedge^3 T^*M)$:
$$
d_MH=0,
\eqno(2.3)
$$
where $d_M$ is the exterior differential of 
$M$. \footnote{}{}\footnote{${}^1$}{The sign convention of the $H$ field
used in this paper is opposite to that of ref. \ref{7}.} 
$H$ measures the so called twisting. The pair $(M,H)$ is called 
a twisted manifold. $(M,H)$ is characterized by the cohomology class 
$[H]\in H^3(M,\Bbb R)$. By definition, $b$ transformation shifts $H$ by the 
exact $3$--form $-d_Mb$:
$$
H'=H-d_Mb.
\eqno(2.4)
$$
So, the cohomology class $[H]$ is invariant.

On a twisted manifold $(M,H)$, there is a natural bilinear pairing 
defined on $C^\infty(TM\oplus T^*M)$ extending the customary Lie pairing on 
$C^\infty(TM)$, called $H$ twisted Courant brackets \ref{40,41}. It is given by the 
expression \xxx
$$
[X+\xi,Y+\eta]_H=[X,Y]+l_X\eta-l_Y\xi-\hbox{$1\over 2$}d_M(i_X\eta-i_Y\xi)+i_Xi_Y H,
\eqno(2.5)
$$
with $X+\xi, Y+\eta\in C^\infty(TM\oplus T^* M)$, where $l_V$ 
denotes Lie derivation with respect to a vector field $V$. 
The pairing is antisymmetric, but it fails to satisfy the 
Jacobi identity. However, remarkably, the Jacobi identity is satisfied 
when restricting to sections $X+\xi, Y+\eta\in C^\infty(L)$,
where $L$ is a subbundle of $TM\oplus T^* M$ isotropic with respect 
to $\langle\,,\rangle$ and involutive (closed) under $[\,,]_H$.
The brackets $[\,,]_H$ are covariant under the action of the diffeomorphism group.
They are also covariant under $b$ transform,
$$
[\exp(b)(X+\xi),\exp(b)(Y+\eta)]_H=\exp(b)[X+\xi,Y+\eta]_{H'},
\eqno(2.6)
$$
where $H'$ is the $b$ transform of $H$ given by (2.4). 

A generalized almost complex structure $\cal J$ is a section of 
$C^\infty(\End(TM\oplus T^* M))$, which is an isometry of 
the metric $\langle\,,\rangle$ and satisfies 
$$
{\cal J}^2=-1.
\eqno(2.7)
$$ 
The pair $(M,{\cal J})$ is called a generalized almost complex manifold.
The group of isometries of $\langle\,,\rangle$ acts on $\cal J$ 
by conjugation. In particular, the $b$ transform of $\cal J$ 
is given by 
$$
{\cal J}'=\exp(-b){\cal J}\exp(b).
\eqno(2.8)
$$ 

If $M$ is equipped with an $H$ field and a generalized almost complex structure 
$\cal J$, the triple $(M,H,{\cal J})$ is called a twisted generalized almost 
complex manifold. The generalized almost complex structure $\cal J$ 
is $H$ integrable if its $\pm\sqrt{-1}$ eigenbundles are involutive with respect to
the twisted Courant brackets $[\,,]_H$. 
\footnote{}{}\footnote{${}^2$}{
The $\pm\sqrt{-1}$ eigenbundles of $\cal J$ are complex and, thus, 
their analysis requires complexifying $TM\oplus T^* M$ leading to 
$(TM\oplus T^* M)\otimes\Bbb C$.}
In that case, $\cal J$ is called an $H$ twisted  generalized complex structure. 
The triple $(M,H,{\cal J})$ is then called a twisted generalized complex manifold.
It can be shown that this condition is equivalent to the vanishing of the appropriate 
$H$ twisted generalized Nijenhuis tensor \xxx
$$
N_H(X+\xi,Y+\eta)=0,
\eqno(2.9)
$$
for all $X+\xi, Y+\eta\in C^\infty(TM\oplus T^* M)$, where
$$
\eqalignno{\vphantom{1\over 2} 
N_H(X+\xi,Y+\eta)&=[X+\xi,Y+\eta]_H+{\cal J}[{\cal J}(X+\xi),Y+\eta]_H&(2.10)\cr
\vphantom{1\over 2}
&\,+{\cal J}[X+\xi, {\cal J}(Y+\eta)]_H-[{\cal J}(X+\xi),{\cal J}(Y+\eta)]_H.&\cr
}
$$
$H$ integrability is preserved by $b$ transformation:
if $\cal J$ is an $H$ twisted generalized complex structure
and $H'$ and ${\cal J}'$ are the $b$ transform of $H$ and $\cal J$, respectively, 
then ${\cal J}'$ is an $H'$ twisted generalized complex structure.

In the untwisted formulation of generalized complex geometry, the $H$ field 
vanishes throughout. To preserve the condition $H=0$, $b$ transformation must 
be restricted. By (2.4), it follows that only closed $b$ fields are allowed, $d_Mb=0$.
This restriction is not necessary in the twisted case, which is the one we deal 
with mostly in this paper.

In practice, it is convenient to decompose a generalized almost complex structure 
$\cal J$ in block matrix form as follows \xxx
$$
{\cal J}=\left(\matrix{J& P&\cr Q& -J^*&\cr}\!\!\!\!\!\!\!\right),
\eqno(2.11)
$$
where $J\in C^\infty(TM\otimes T^* M)$, $P\in C^\infty(\wedge^2 TM)$, 
$Q\in C^\infty(\wedge^2 T^* M)$ and express all the properties of $\cal J$,
in particular its integrability, in terms of the blocks $J$, $P$, $Q$. 

For later use, we write in explicit tensor notation the conditions obeyed by the 
fields $H$, $J$, $P$, $Q$.

An $H$ field satisfies the closedness equation
$$
\partial_aH_{bcd}-\partial_bH_{acd}+\partial_cH_{abd}-\partial_dH_{abc}=0,
\eqno(2.12)
$$
(cf. eq. (2.3)). Under $b$ transform, we have
$$ 
H'{}_{abc}=H_{abc}-\big(\partial_ab_{bc}+\partial_bb_{ca}+\partial_cb_{ab}\big),
\eqno(2.13)
$$
(cf. eq. (2.4)).

If $\cal J$ is a generalized almost complex structure, the tensors 
$J$, $P$, $Q$ satisfy 
$$
\eqalignno{
\vphantom{1\over 2}
&P^{ab}+P^{ba}=0,&(2.14a)\cr
\vphantom{1\over 2}
&Q_{ab}+Q_{ba}=0,&(2.14b)\cr
\vphantom{1\over 2}
&J^a{}_cJ^c{}_b+P^{ac}Q_{cb}+\delta^a{}_b=0,&(2.15a)\cr
\vphantom{1\over 2}
&J^a{}_cP^{cb}+J^b{}_cP^{ca}=0,&(2.15b)\cr
\vphantom{1\over 2}
&Q_{ac}J^c{}_b+Q_{bc}J^c{}_a=0,&(2.15c)\cr
}
$$
on account of (2.7). Upon using (2.2), (2.8), we find that, under $b$ transform,
$$
\eqalignno{
\vphantom{1\over 2} 
&P'{}^{ab}=P^{ab},&(2.16a)\cr
\vphantom{1\over 2} 
&J'{}^a{}_b=J^a{}_b-P^{ac}b_{cb},&(2.16b)\cr
\vphantom{1\over 2} 
&Q'{}_{ab}=Q_{ab}+b_{ac}J^c{}_b-b_{bc}J^c{}_a+P^{cd}b_{ca}b_{db}.&(2.16c)\cr
}
$$

The $H$ integrability condition (2.9)
of a generalized almost complex structure $\cal J$ 
can be cast in the form of a set of four tensorial equations 
$$
\eqalignno{
\vphantom{1\over 2} 
&A_H{}^{abc}=0,&(2.17a)\cr
\vphantom{1\over 2} 
&B_H{}_a{}^{bc}=0,&(2.17b)\cr
\vphantom{1\over 2} 
&C_H{}_{ab}{}^c=0,&(2.17c)\cr
\vphantom{1\over 2}
&D_H{}_{abc}=0,&(2.17d)\cr
}
$$
where $A_H$, $B_H$, $C_H$, $D_H$ are the tensors defined by  
$$
\eqalignno{\vphantom{1\over 2} 
A_H{}^{abc}&=P^{ad}\partial_dP^{bc}+P^{bd}\partial_dP^{ca}+P^{cd}\partial_dP^{ab},
&(2.18a)\cr
\vphantom{1\over 2}
B_H{}_a{}^{bc}&=J^d{}_a\partial_dP^{bc}
+P^{bd}(\partial_aJ^c{}_d-\partial_d J^c{}_a)
-P^{cd}(\partial_aJ^b{}_d -\partial_dJ^b{}_a)&(2.18b)\cr
\vphantom{1\over 2}
&\,-\partial_a(J^b{}_dP^{dc})+P^{bd}P^{ce}H_{ade},&\cr
\vphantom{1\over 2}
C_H{}_{ab}{}^c&=J^d{}_a\partial_dJ^c{}_b-J^d{}_b\partial_dJ^c{}_a
-J^c{}_d\partial_aJ^d{}_b+J^c{}_d\partial_bJ^d{}_a&(2.18c)\cr
\vphantom{1\over 2}
&\,+P^{cd}(\partial_dQ_{ab}+\partial_aQ_{bd}+\partial_bQ_{da})
-J^d{}_aP^{ce}H_{bde}+J^d{}_bP^{ce}H_{ade},&\cr
}
$$
$$
\eqalignno{
\vphantom{1\over 2}
D_H{}_{abc}&=J^d{}_a(\partial_dQ_{bc}+\partial_bQ_{cd}+\partial_cQ_{db})
+J^d{}_b(\partial_dQ_{ca}+\partial_cQ_{ad}+\partial_aQ_{dc})&(2.18d)\cr
\vphantom{1\over 2}
&+J^d{}_c(\partial_dQ_{ab}+\partial_aQ_{bd}+\partial_bQ_{da})
-\partial_a(Q_{bd}J^d{}_c)-\partial_b(Q_{cd}J^d{}_a)-\partial_c(Q_{ad}J^d{}_b)&\cr
\vphantom{1\over 2}
&\,-H_{abc}+J^d{}_aJ^e{}_bH_{cde}+J^d{}_bJ^e{}_cH_{ade}+J^d{}_cJ^e{}_aH_{bde}.&\cr
}
$$
The above expressions were first derived in a different but equivalent form in 
\ref{25} and subsequently in the form given here in \ref{31}. 

One of the most interesting features of generalized complex geometry is its 
capability for a unified treatment of complex and symplectic geometry, as we  
show below. However, as noticed by Hitchin, these geometries
do not exhaust the scope of generalized complex geometry. 
In fact, there are manifolds which cannot support 
any complex or symplectic structures, but do admit generalized 
complex structures \ref{6}. These facts explain the reason why Hitchin's construction 
is interesting and worthwhile pursuing and not simply an elegant 
repackaging of known notions.

Complex geometry can be formulated in terms of generalized almost 
complex structures $\cal J$ of the form \xxx
$$
{\cal J}=\left(\matrix{J & 0 &\cr
                       0 & -J^*&\cr}\!\!\!\!\!\!\right),
\eqno(2.19)
$$
where $J$ is an ordinary almost complex structures i.e $J^a{}_cJ^c{}_b
=-\delta^a{}_b$. $\cal J$ satisfies (2.17{\it a--d}) if $J$, $H$
satisfy \xxx
$$
J^d{}_a\partial_dJ^c{}_b-J^d{}_b\partial_dJ^c{}_a
-J^c{}_d\partial_aJ^d{}_b+J^c{}_d\partial_bJ^d{}_a=0,
\eqno(2.20)
$$
$$
H_{abc}-J^d{}_aJ^e{}_bH_{cde}-J^d{}_bJ^e{}_cH_{ade}-J^d{}_cJ^e{}_aH_{bde}=0.
\eqno(2.21)
$$
Eq. (2.20) is nothing but the statement of the vanishing of the Nijenhuis 
tensor of $J$ and, so, it implies that $J$ is a complex structure.
Eq. (2.21) states in turn the $3$--form $H$ is of type $(2,1)+(1,2)$
with respect to $J$. 

Similarly, symplectic geometry can be formulated in terms of generalized almost 
complex structures $\cal J$ of the form \xxx 
$$
{\cal J} = \left (\matrix{0 & -Q^{-1} &\cr
                          Q & 0&\cr}\!\!\!\!\!\!\right),
\eqno(2.22)
$$
where $Q$ is pointwise non singular a $2$--form. $\cal J$ satisfies (2.17{\it a--d}) 
if $Q$, $H$ satisfy 
$$
\partial_aQ_{bc}+\partial_bQ_{ca}+\partial_cQ_{ab}=0,
\eqno(2.23)
$$
$$
H_{abc}=0.
\eqno(2.24)
$$
Eq. (2.23) states that the $2$--form $Q$ is closed and, so, it implies that $Q$ is a 
symplectic structure. By eq. (2.24), the $H$ field necessarily
vanishes in the symplectic case. 

In the Hitchin sigma model studied in \ref{31}, the action $S$ 
contains a topological Wess-Zumino term defined up to the periods of 
the closed $3$--form $H$. In the quantum path integral, so, the weight 
$\exp(\sqrt{-1}S)$ is unambiguously defined only if $H/2\pi$ has integer 
periods. Therefore, the cohomology class $[H/2\pi]$ belongs to the image of 
$H^3(M,\Bbb Z)$ in $H^3(M,\Bbb R)$. This case is particular important for 
its relation to gerbes. \footnote{}{}\footnote{${}^3$}{
See \ref{42} for background material about this topic.}
In this context, the $b$ tansforms with $b$ a closed $2$--form such that 
$[b/2\pi]$ is contained in the image of $H^2(M,\Bbb Z)$ in $H^2(M,\Bbb R)$ 
represent the gerbe generalization of gauge transformations.

In \ref{43}, it was shown that a sigma model on a manifold $M$ with
NSNS background $H$ has $(2,2)$ supersymmetry if the twisted manifold
$(M,H)$ is ``Kaehler with torsion''. This means that $M$ is equipped with a 
Riemannian metric $g$ and two generally different complex structures $J_\pm$ such 
that $g$ is Hermitian with respect to both $J_\pm$ and that $J_\pm$ are parallel 
with respect to two different metric connections $\nabla_\pm$ with torsion 
proportional to $\pm H$. The presence of torsion implies that the 
geometry is not Kaehler. As shown in \ref{7}, these geometrical data can be 
assembled in a pair of commuting $H$ twisted generalized complex structures 
${\cal J}_i$, $i=1$, $2$, describing a generalized Kaehler geometry.

The $(2,2)$ supersymmetric sigma model has been studied mostly for 
$H=0$ and $J_+=J_-$, when $M$ is a true Kaehler manifold.  
In this case, the generalized complex structures ${\cal J}_1$
${\cal J}_2$ are of the special form (2.19), (2.22) and encode the
complex and symplectic geometry of $M$, respectively. 
The associated A and B topological sigma models depend on only one 
of these, depending on the topological twisting used \ref{8}: 
the A model depends only on ${\cal J}_2$,  the B model only on ${\cal J}_1$.
A topological sigma model for generalized Kaehler geometry has been proposed 
in \ref{13}. 
The Hitchin sigma model \ref{31}, which is the main topic of this paper, 
can be defined for a general twisted generalized complex structure
$\cal J$ not necessarily of the form of those appearing in generalized Kaehler 
geometry. 

\titlebf{3. Generalized complex submanifolds and branes}

\par
According to \ref{7}, a generalized submanifold of a twisted manifold $(M,H)$ 
(cf. sect. 2) is a 
pair $(W,F)$, where $W$ is a submanifold of $M$ and $F\in C^\infty(\wedge^2T^*W)$ 
is such that 
$$
\iota_W{}^*H=d_WF,
\eqno(3.1)
$$
where $\iota_W:W\rightarrow M$ is the natural injection and $d_W$ is the differential
of $W$. Note that, by (2.3) and (3.1), the pair $(H,F)$ is a representative of a 
relative cohomology class in $H^3(M,W,\Bbb R)$. By (2.4), (3.1), 
$F$ transforms non trivially under a $b$ transform, viz
$$
F'=F-\iota_W{}^*b.
\eqno(3.2)
$$
where $b\in C^\infty(\wedge^2 T^* M)$.
In this way, if $(W,F)$ is a generalized submanifold of $(M,H)$, 
then $(W,F')$ is a generalized submanifold of $(M,H')$, 
where $H'$, $F'$ are the $b$ transforms of $H$, $F$ given by (2.4), (3.2).

The generalized tangent bundle ${\cal T}^FW$ of $(W,F)$ is the subbundle of 
$TM\oplus T^*M|_W$ spanned by the restriction to $W$ of those sections 
$X+\xi\in C^\infty(TM\oplus T^* M)$ such that $X|_W\in C^\infty(TW)$ and that \xxx 
$$
\iota_W{}^*\xi=i_{X|_W}F.
\eqno(3.3)
$$
(Here, $TW$ is viewed as a subbundle of $TM|_W$.)
It is easy to see that $b$ transform acts naturally on 
${\cal T}^FW$. Indeed, one has $\exp(b){\cal T}^{F'}W={\cal T}^FW$.
In fact, this is the main reason why ${\cal T}^FW$ 
is defined the way indicated above. 

Suppose $(M,H,{\cal J})$ is a twisted generalized almost complex manifold
(cf. sect. 2). A generalized submanifold $(W,F)$ of 
$(M,H)$ is a generalized almost complex submanifold 
of $(M,H,{\cal J})$, if ${\cal T}^FW$ is stable under the action of $\cal J$
seen here as a section of $C^\infty(\End(TM\oplus T^* M))$. When $\cal J$ is $H$ 
integrable and, therefore, $(M,H,{\cal J})$ is a twisted generalized complex manifold, 
we call $(W,F)$ a generalized complex submanifold of $(M,H,{\cal J})$.
\footnote{}{}\footnote{${}^4$}{A generalization of Gualtieri's definition of 
generalized complex submanifold can be found in ref. \ref{44}.}
This notion is covariant under $b$ transformation:
if $(W,F)$ is a generalized almost complex submanifold of $(M,H,{\cal J})$, 
then $(W,F')$ is a generalized almost complex submanifold of $(M,H',{\cal J}')$, 
where $H'$, ${\cal J}'$, $F'$ are the $b$ transforms of $H$, $\cal J$, $F$ 
given by (2.4), (2.8), (3.2).

As is well known, at each point of $W$, there are coordinates of $M$
$t^a=(t^i,t^\rho)$ such that, locally, the submanifold $W$ is described by the 
equation $t^\rho=0$ and coordinatized by the $t^i$. We call such coordinates
adapted. Here, middle Latin indices $i,~j,~\ldots$ take the values $1,~\ldots, 
\dim W$, while late Greek indices $\rho,~\sigma,~\ldots$ take the values
$1,~\ldots, d-\dim W$. The abstract geometrical notions outlined above are 
conveniently expressed in terms of adapted coordinates. The relations so obtained 
are adapted covariant, that is they have the same mathematical form for all choices 
of adapted coordinates. The tensor components $H_{abc}$, $P^{ab}$, $J^a{}_b$
$Q_{ab}$ of the $H$ field and the blocks $P$, $J$, $Q$ 
of the generalized almost complex 
structure $\cal J$ (cf. eq. (2.11) entering in them 
are tacitly assumed to be restricted to $W$ to avoid the cumbersome 
repetition of $\big|_W$. Since the $F$ field 
is defined only on $W$ anyway, no restriction of 
the components $F_{ij}$ is involved. 

Relation (3.1), connecting $H$ and $F$ reads simply
$$
H_{ijk}=\partial_i F_{jk}+\partial_j F_{ki}+\partial_k F_{ij}
\eqno(3.4)
$$
at $W$. Under a $b$ transform, one has \xxx
$$
F'{}_{ij}=F_{ij}-b_{ij},
\eqno(3.5)
$$
as follows from (3.2).

If $X+\xi\in C^\infty(TM\oplus T^* M)$ restricts to a section of the generalized 
tangent space ${\cal T}^FW$, then one has \xxx 
$$
\eqalignno{\vphantom{1\over 2} 
&X^\rho=0,
&(3.6a)\cr
\vphantom{1\over 2}
&\xi_i+F_{ij}X^j=0&(3.6b)\cr
}
$$ 
at $W$, as follows from (3.3).

If $(M,H,{\cal J})$ is a twisted generalized almost complex manifold and 
$(W,F)$ is a generalized almost complex submanifold of $(M,H,{\cal J})$,
then, in the block representation (2.11) of $\cal J$, one has \xxx
$$
\eqalignno{\vphantom{1\over 2} 
&P^{\rho\sigma}=0,
&(3.7a)\cr
\vphantom{1\over 2}
&J^\rho{}_i-P^{\rho j}F_{ji}=0,&(3.7b)\cr
\vphantom{1\over 2}
&Q_{ij}-J^k{}_iF_{jk}+J^k{}_jF_{ik}+P^{kl} F_{ki}F_{lj}=0&(3.7c)\cr 
}
$$ 
at $W$. These relations follow from (3.6{\it a, b}) and imposing the stability of 
${\cal T}^FW$ under the action of $\cal J$.
It is straightforward to verify that these conditions are compatible with 
$H$ integrability conditions (2.17{\it a--d}). 

Suppose that $\cal J$ is an $H$ twisted generalized complex structure of the form 
(2.19). Then, $J$, $H$  satisfy (2.20), (2.21) and, so, $J$ is a complex 
structure and $H$ is $(2,1)+(1,2)$ $3$--form.
In this case, eqs. (3.7{\it a--c}) become simply
$$
\eqalignno{
\vphantom{1\over 2}
&J^\rho{}_i=0,&(3.8a)\cr
\vphantom{1\over 2}
&J^k{}_iF_{kj}+J^k{}_jF_{ik}=0&(3.8b)\cr
}
$$ 
at $W$. Eq. (2.20), (3.8{\it a}) entails that the $J^i{}_j$ are the components
of a complex structure $J_W$ on $W$, which is therefore a complex submanifold of $M$. 
Eq. (3.8{\it b}) entails in turn that $F$ is (1,1) $2$--form of $W$. 

Suppose that $\cal J$ is an $H$ twisted generalized complex structure of the form 
(2.22). Then, $Q$, $H$  satisfy (2.23), (2.24) and, so, $Q$ is a symplectic structure
and $H$ vanishes. Then, eqs. (3.7{\it a--c}) become simply
$$
\eqalignno{\vphantom{1\over 2} 
&Q^{-1\rho\sigma}=0,
&(3.9a)\cr
\vphantom{1\over 2}
&Q^{-1\rho j}F_{ji}=0,&(3.9b)\cr
\vphantom{1\over 2}
&Q_{ij}-Q^{-1kl}F_{ki}F_{lj}=0&(3.9c)\cr
}
$$ 
at $W$. Further, as $H=0$, (3.4) furnishes \xxx
$$
\partial_i F_{jk}+\partial_j F_{ki}+\partial_k F_{ij}=0.
\eqno(3.10)
$$
Eq. (3.9{\it a}) entails that $Q^{-1}$ maps the conormal bundle $N^*W$ into 
the tangent bundle $TW$ of $W$. This means that, by definition, that $W$ is 
coisotropic. Condition (3.10) entails that $F$ is a closed $2$--form.
The remaining conditions
(3.9{\it b, c}) do not have any simple geometrical interpretation. (See however 
\ref{7} for an attempt in this direction.) The interpretation of the geometry 
of $W$ simplifies if one adds by hand the condition 
$$
Q_{ij}=0
\eqno(3.11)
$$
at $W$,
in virtue of which (3.9{\it b, c}) are automatically fulfilled, as is easy to see. 
Eq. (3.11) entails that $Q$ maps the tangent bundle $TW$ into the conormal bundle 
$N^*W$ of $W$. 
Hence, when (3.7{\it a}) and (3.11) are satisfied, the submanifold $W$ is 
simultaneously isotropic and coisotropic, hence Lagrangian.
This means that, by definition, that $W$ is isotropic. (3.11) further entails the 
vanishing of the $F$ field \xxx
$$
F_{ij}=0.
\eqno(3.12)
$$

In this paper, we shall provide fresh evidence in favor of the claim 
that, in a consistent sigma model on a twisted generalized complex manifold
$(M,H,{\cal J})$, branes should be generalized complex submanifolds $(W,F)$ 
\ref{10,15,23}. 
We note that the formalism expounded above is suitable only for the description of 
non coincident branes. In the case, often considered in string theory, of stacks of 
overlapping branes, one would need a non Abelian generalization of the above 
construction which is not yet available \ref{19}.

In the brane Hitchin sigma model studied in this paper, branes are generic 
generalized complex submanifolds. So, we call them generalized branes.

The action $S_W$ of the model 
contains a topological Wess-Zumino term defined up to the relative periods of 
the closed relative $3$--form $(H,F)$. In the quantum path integral, so, the weight 
$\exp(\sqrt{-1}S_W)$ is unambiguously defined only if $(H/2\pi,F/2\pi)$ has integer 
relative periods. Therefore, the relative cohomology class $[(H/2\pi,F/2\pi)]$ 
belongs to the image of $H^3(M,W,\Bbb Z)$ in $H^3(M,W,\Bbb R)$. This indicates that 
generalized branes support relative gerbes with connection and curving \ref{42}.

In the A and B topological sigma models with branes, the $H$ field vanishes
and the branes support Abelian gauge fields of field strength $F$. 
The ends of open strings lie on the branes and carry Chan--Paton point charges, 
which couple to gauge fields of the branes. 

As is well known, the consistent interaction of a point charge with a gauge field
requires that the gauge field strength has quantized fluxes. 
Since the open strings carry Chan--Paton point charges coupling to gauge fields 
of the branes, the brane gauge curvature $F/2\pi$ has integer periods.
It follows that the A-- and B--branes always support line bundles 
with connections \ref{19}. 

Since open strings are involved, 
the world sheets have non empty boundaries and the world sheet fields 
obey boundary conditions, whose nature depends on the topological twisting 
and restricts the type of submanifolds of the Kaehler target space 
branes can be \ref{21}.
In the A model, in its simplest form, the boundary conditions are such that 
the complex structure exchanges normal and tangent directions 
to the brane and, for this reason, an A--brane must be a Lagrangian 
submanifold with respect to the Kaehler symplectic structure. 
In the B model, conversely, the boundary conditions make  
the complex structure preserve normal and tangent directions 
and, so, a B--brane must be a complex submanifold with respect to the Kaehler 
complex structure. 

Further restrictions arise from requirement that the coupling
of the open string Chan--Paton charges to the brane gauge fields  
is invariant under the nilpotent topological charge \ref{19}.
In A model, this implies that the field strength $F$ vanishes, so that the 
underlying line bundle is flat. In the B model, conversely, this makes 
$F$ be a $(1,1)$ $2$--form and the underlying line bundle holomorphic.  

Other restrictions follow from the requirement of ghost number anomaly cancellation. 
In the A model, this requires that the branes are graded Lagrangian submanifolds 
of vanishing Maslov class \ref{20}. 

To summarize, the above qualitative discussion indicates that an A--brane is a
Lagrangian submanifold carrying a flat line bundle with flat connection and 
that a B--brane is a complex submanifold carrying a holomorphic line bundle
with connection with $(1,1)$ curvature.
The careful analysis of Kapustin and Orlov \ref{22} shows that, in the A model,  
a class of coisotropic submanifolds carrying non flat line bundles
can also serve as A--branes, at least at the classical level. 
(See also the recent proposal of ref. \ref{15}.)

The above brief review of how branes show up in the A and B topological sigma models
fits quite well Gualtieri's formalism of generalized complex submanifolds illustrated 
above. Indeed, A-- and B--branes are generalized complex submanifolds of the types 
described in the paragraphs of eqs. (3.9{\it a}) and (3.8{\it a}), respectively,
and, thus, also particular examples of generalized branes

\titlebf{4. 2--dimensional de Rham superfields}

In general, the fields of a 2--dimensional field theory are differential 
forms on a oriented $2$--dimensional manifold $\Sigma$. They can be viewed 
as elements of the space $\Fun(\Pi T\Sigma)$ of functions on the parity 
reversed tangent bundle $\Pi T\Sigma$ of $\Sigma$, which we shall call 
de Rham superfields \ref{31,35}. 
More explicitly, we associate with the coordinates $z^\alpha$ of 
$\Sigma$ Grassmann odd partners $\zeta^\alpha$ 
with \xxx
$$
\deg z^\alpha=0, \qquad \deg\zeta^\alpha=1.\vphantom{\Big[}
\eqno(4.1)
$$
$\zeta^\alpha$ transforms as the differential of $z^\alpha$ under coordinate changes. 
A de Rham superfield $\psi(z,\zeta)$ is a triplet 
formed by a $0$--, $1$--, $2$--form field $\psi^{(0)}(z)$, 
$\psi^{(1)}{}_\alpha(z)$, $\psi^{(2)}{}_{\alpha\beta}(z)$ set as
$$
\psi(z,\zeta)=\psi^{(0)}(z)+\zeta^\alpha\psi^{(1)}{}_\alpha(z)
+\hbox{$1\over 2$}\zeta^\alpha\zeta^\beta\psi^{(2)}{}_{\alpha\beta}(z).
\eqno(4.2)
$$
The forms $\psi^{(0)}$, $\psi^{(1)}$, $\psi^{(2)}$ are called the 
de Rham components of $\psi$. 

$\Pi T\Sigma$ is endowed with a natural differential $d$ defined by 
$$
dz^\alpha=\zeta^\alpha,\qquad d\zeta^\alpha=0.\vphantom{\Big[}
\eqno(4.3)
$$
In this way, the exterior differential $d$ of $\Sigma$ 
can be identified with the operator 
$$
d=\zeta^\alpha{\partial\over\partial z^\alpha}.
\eqno(4.4)
$$

The coordinate invariant integration measure of $\Pi T\Sigma$ is 
$$
\mu={\rm d}z^1{\rm d}z^2{\rm d}\zeta^1{\rm d}\zeta^2.
\eqno(4.5)
$$
Any de Rham superfield $\psi$ can be integrated on $\Pi T\Sigma$ according to
the prescription
$$
\int_{\Pi T\Sigma}\mu\,\psi=\int_\Sigma\hbox{$1\over 2$}
{\rm d}z^\alpha {\rm d}z^\beta\psi^{(2)}{}_{\alpha\beta}(z).
\eqno(4.6)
$$

Similarly, fields on the boundary $\partial\Sigma$ can be viewed as elements 
of the space $\Fun(\Pi T$ $\partial\Sigma)$ of functions on the parity 
reversed tangent bundle $\Pi T\partial\Sigma$ of $\partial\Sigma$, which we shall 
call boundary de Rham superfields. Again, we associate with the coordinates $s$ of 
$\partial\Sigma$ Grassmann odd partners $\varsigma$ 
with \xxx
$$
\deg s=0, \qquad \deg\varsigma=1.\vphantom{\Big[}
\eqno(4.7)
$$
$\varsigma$ transforms as the differential of $s$ under coordinate changes. 
A generic boundary de Rham superfield $\chi(s,\varsigma)$ is a doublet 
formed by a $0$--, $1$--form field $\chi^{(0)}(s)$, $\chi^{(1)}{}_s(s)$, 
organized as
$$
\chi(s,\varsigma)=\chi^{(0)}(s)+\varsigma\chi^{(1)}{}_s(s).
\eqno(4.8)
$$
The forms $\chi^{(0)}$, $\chi^{(1)}$ are called the boundary  de Rham components 
of $\chi$. 

$\Pi T\partial\Sigma$ is endowed with a natural differential $d_\partial$ defined by 
$$
d_\partial s=\varsigma,\qquad d_\partial\varsigma=0.\vphantom{\Big[}
\eqno(4.9)
$$
The exterior differential $d_\partial$ of $\partial\Sigma$ 
can be identified in this way with the operator 
$$
d_\partial=\varsigma\,{d\over d s}.
\eqno(4.10)
$$

The coordinate invariant integration measure of $\Pi T\partial\Sigma$ is 
$$
\mu_\partial={\rm d}s{\rm d}\varsigma.
\eqno(4.11)
$$
Any de Rham superfield $\chi$ can be integrated on $\Pi T\partial\Sigma$ according to
the prescription
$$
\oint_{\Pi T\partial\Sigma}\mu_\partial\,\chi=\oint_{\partial\Sigma}
{\rm d}s \chi^{(1)}{}_s(s).
\eqno(4.12)
$$

Let $\iota_\partial:\partial\Sigma\rightarrow\Sigma$ be the natural injection. 
If $\psi$ is a de Rham superfield of $\Sigma$, then 
$$
\iota_\partial{}^*\psi(s,\varsigma)=\psi^{(0)}(\iota_\partial(s))+
\varsigma\,{d\iota_\partial{}^\alpha\over ds}(s)\psi^{(1)}{}_\alpha(\iota_\partial(s))
\eqno(4.13)
$$
is a boundary de Rham superfield, the pull--back of $\psi$.
By Stokes' theorem, 
$$
\int_{\Pi T\Sigma}\mu\, d\psi
=\oint_{\Pi T\partial\Sigma}\mu_\partial \iota_\partial{}^*\psi.
\eqno(4.14)
$$
Often, for the sake of simplicity, we shall write 
$\psi$ rather than $\iota_\partial{}^*\psi$ in the right hand side.
Similarly, by the boundary Stokes' theorem,  
if $\chi$ is a boundary de Rham superfield,
$$
\oint_{\Pi T\partial\Sigma}\mu_\partial d_\partial\chi=0.
\eqno(4.15)
$$

It is possible to define functional derivatives of functionals of de Rham superfields.
Let $\psi$ be a de Rham superfield and let $F(\psi)$ be a functional of $\psi$.
We define the left/right functional derivative superfields 
$\delta_{l,r} F(\psi)/\delta\psi$ as follows. 
Let $\sigma$ be a superfield of the same properties as $\psi$.
Then, 
$$
{d\over dt}F(\psi+t\sigma)\Big|_{t=0}
=\int_{\Pi T\Sigma}\mu\,\sigma{\delta_l F(\psi)\over\delta\psi}
=\int_{\Pi T\Sigma}\mu\,{\delta_r F(\psi)\over\delta\psi}\sigma.
\eqno(4.16)
$$
To write $dF(\psi+t\sigma)/dt\big|_{t=0}$ as an integral on $\Pi T\Sigma$ as 
above, repeated applications of Stokes' theorem are required. This yields 
boundary terms, which must cancel out. In general, this is possible 
only if suitable boundary conditions are imposed in the superfields.

In the applications below, the components of the relevant de Rham superfields
carry, besides the form degree, also a ghost degree. We shall limit ourselves 
to homogeneous superfields. A de Rham superfield $\psi$ is said homogeneous 
if the sum of the form and ghost degree is the same for all its components 
$\psi^{(0)}$, $\psi^{(1)}$, $\psi^{(2)}$ of $\psi$. 
The common value of that sum is called the (total) degree $\deg\psi$ of $\psi$. 
It is easy to see that the differential operator $d$ and the
integration operator $\int_{\Pi T\Sigma}\mu$ carry 
degree $1$ and $-2$, respectively. 
Also, if $F(\psi)$ is a functional of a superfield $\psi$, then
$\deg\delta_{l,r} F(\psi)/\delta\psi=\deg F-\deg \psi+2$.
Similar considerations hold for the boundary de Rham superfields.
In this case, the differential operator $d_\partial$ and the
integration operator $\int_{\Pi T\partial\Sigma}\mu_\partial$ carry 
degree $1$ and $-1$, respectively.

\titlebf{5. The Hitchin sigma model in the presence of branes}

In this section, we shall first review the formulation of the Hitchin sigma 
model in the absence of branes \ref{31}. Subsequently, we shall show how to 
incorporate branes into it. 

The Hitchin sigma model is closely related to the standard Poisson sigma model
\ref{33--34}, of which it has the same field content. The approach used here is 
based on the Batalin--Vilkovisky quantization scheme \ref{35--37}. To make the 
treatment as simple and transparent as possible, we shall use the convenient 
de Rham superfield formalism (cf. sect. 4) following the original work of 
Cattaneo and Felder in \ref{35} (see also \ref{45--47}).
We shall limit ourselves to the lowest order in perturbation theory, since the 
constraints on target space geometry following from the Batalin--Vilkovisky 
classical master equation lead directly to Hitchin's generalized complex geometry. 
Quantum corrections will presumably yield a deformation of the latter, whose study 
is beyond the scope of this paper. We will not attempt the gauge fixing of the 
field theory, which, at any rate, is expected to be essentially identical to that 
of the ordinary Poisson sigma model as described in \ref{35,45}.
We shall consider directly the twisted version of the Hitchin sigma model. 
The umtwisted version can be readily obtained by setting the $H$ field to zero.

When branes are absent, the target space of the Hitchin sigma model is a twisted 
manifold $(M,H)$ (cf. sect. 2). The base space is a closed oriented surface $\Sigma$. 

The basic fields of the standard Hitchin sigma model are a degree $0$
superembedding $x\in\Gamma(\Pi T\Sigma, M)$ and a degree $1$
supersection $y\in\Gamma(\Pi T\Sigma, x^*\Pi T^*M)$, where $\Pi$ is 
the parity inversion operator. With respect to each local coordinate $t^a$ 
of $M$, $x$, $y$ are given as de Rham superfields $x^a$, $y_a$. 
Under a change of coordinates, these transform as 
$$
\eqalignno{\vphantom{1\over 2}
x'^a&=t'^a\circ t^{-1}(x)&(5.1)\cr
\vphantom{1\over 2}
y'{}_a&=\partial'{}_a t^b\circ t^{-1}(x)y_b.&(5.2)\cr
}
$$
The resulting transformation rules of the de Rham components of $x^a(z,\zeta)$, 
$y_a(z,\zeta)$ are obtainable by expanding these relations in powers of 
$\zeta^\alpha$.

The Batalin--Vilkovisky odd symplectic form is 
$$
\Omega=\int_{\Pi T\Sigma}\mu\,\Big[\delta x^a\delta y_a
+\hbox{$1\over 2$}H_{abc}(x)\delta x^a dx^b \delta x^c\Big].
\eqno(5.3)
$$
$\Omega$ is not of the canonical form when $H\not=0$. Hence, $x^a$, $y_a$ 
are not canonical fields/anti\-fields. However, $\Omega$ is a
closed functional form, $\delta\Omega=0$. In this way, one can define 
antibrackets $(\,,)$ in standard fashion. The resulting expression is 
$$
(F,G)=\int_{\Pi T\Sigma}\mu\,\bigg[
{\delta_r F\over\delta x^a}{\delta_l G\over\delta y_a}-
{\delta_r F\over\delta y_a}{\delta_l G\over\delta x^a}
-H_{abc}(x){\delta_r F\over\delta y_a} 
dx^b {\delta_l G\over\delta y_c}\bigg],
\eqno(5.4)
$$
for any two functionals $F$, $G$ of $x^a$, $y_a$.

In the Hitchin sigma model, the target space geometry is specified by a 
generalized almost complex structure $\cal J$ (cf.sect. 2). 
In the representation (2.11), the action is 
$$
\eqalignno{\vphantom{1\over 2}
S&=\int_{\Pi T\Sigma}\mu\,\Big[y_adx^a+\hbox{$1\over 2$}P^{ab}(x)y_ay_b
+\hbox{$1\over 2$}Q_{ab}(x)dx^adx^b+J^a{}_b(x)y_adx^b\Big]&(5.5)\cr
\vphantom{1\over 2}
&-2\int_\Gamma \bar x^{(0)*}H.&\cr
}
$$
Here, $\Gamma$ is a $3$--fold such that \xxx
$$
\partial \Gamma=\Sigma.
\eqno(5.6)
$$
$ \bar x^{(0)}:\Gamma\rightarrow M$ is an embedding such that $ \bar x^{(0)}|_\Sigma$
equals the lowest degree $0$ component $x^{(0)}$ of the superembedding $x$.
The last $H$ dependent term is a Wess--Zumino like term. 
A similar term was added to the action of the standard Poisson sigma model 
in ref. \ref{48}. Its value depends on the the choice of 
$\bar x^{(0)}$. In the quantum theory, in order to have a 
well defined weight $\exp(\sqrt{-1}S)$ in the path integral, 
it is necessary to require that $H/2\pi$ has integer 
periods, so that the cohomology class $[H/2\pi]\in H^3(M,\Bbb R)$ 
belongs to the image of $H^3(M,\Bbb Z)$ in $H^3(M,\Bbb R)$ (cf. sect. 2).
We note that the above definition of the topological term works only if
$x^{(0)}(\Sigma)$ is a boundary in $M$. If one wants to extend the definition 
to the general case where $x^{(0)}(\Sigma)$ is a cycle of $M$, the theory of 
Cheeger--Simons differential characters is required \ref{49,50}. 

A straightforward computation furnishes 
$$
\eqalignno{\vphantom{1\over 2} 
(S,S)=
2\int_{\Pi T\Sigma}\mu\,\Big[&-\hbox{$1\over 6$}A_H{}^{abc}(x)y_ay_by_c
+\hbox{$1\over 2$}B_H{}_a{}^{bc}(x)dx^ay_by_c&(5.7)\cr
\vphantom{1\over 2}
&-\hbox{$1\over 2$}C_H{}_{ab}{}^c(x)dx^adx^by_c
+\hbox{$1\over 6$}D_H{}_{abc}(x)dx^adx^bdx^c&\cr
\vphantom{1\over 2}
&-\hbox{$1\over 3$}H_{abc}(x)dx^adx^bdx^c\Big],
&\cr}
$$
where the tensors $A_H$, $B_H$, $C_H$, $D_H$ are given by (2.18{\it a--d}). Hence, 
$S$ satisfies the classical Batalin--Vilkovisky master equation 
$$
(S,S)=0
\eqno(5.8)
$$
if (2.17{\it a--d}) hold, i.e. when $\cal J$ is an $H$ twisted generalized complex 
structure and, so, $(M,H,{\cal J})$ a twisted generalized complex manifold. 
(Recall that $dx^adx^bdx^c=0$ on $\Pi T\Sigma$.)
This shows that {\it there is a non trivial connection between generalized complex 
geometry and quantization \`a la Batalin--Vilkovisky of the sigma model} \ref{31}.
(2.17{\it a--d}) are sufficient but not necessary conditions for the fulfillment 
of the master equation (5.8). 

The Batalin--Vilkovisky variations $\delta_{BV}x^a$, 
$\delta_{BV}y_a$ are given by 
$$
\eqalignno{\vphantom{1\over 2} 
\delta_{BV} x^a&=(S,x^a),&(5.9a)\cr
\vphantom{1\over 2}
\delta_{BV} y_a&=(S,y_a).&(5.9b)\cr
}
$$
Using (5.4), we can then derive the expressions of the 
Batalin--Vilkovisky variations $\delta_{BV} x^a$, $\delta_{BV} y_a$. The result is 
$$
\eqalignno{
\vphantom{1\over 2} 
\delta_{BV} x^a&=dx^a+P^{ab}(x)y_b+J^a{}_b(x)dx^b,&(5.10a)\cr
\vphantom{1\over 2}
\delta_{BV} y_a&=dy_a+\hbox{$1\over 2$}\partial_aP^{bc}(x)y_by_c
+\hbox{$1\over 2$}(\partial_aQ_{bc}+\partial_bQ_{ca}+\partial_cQ_{ab})(x)dx^bdx^c&
(5.10b)\cr
\vphantom{1\over 2}
&+(\partial_aJ^b{}_c-\partial_cJ^b{}_a)(x)y_bdx^c+J^b{}_a(x)dy_b\cr
\vphantom{1\over 2}
&+\hbox{$1\over 2$}(H_{abd}J^d{}_c-H_{acd}J^d{}_b)(x)dx^bdx^c
+H_{adc}P^{db}(x)y_bdx^c.&\cr
}
$$
As well-known \ref{36,37}, the Batalin--Vilkovisky variation operator 
$\delta_{BV}$ is nilpotent, 
$$
\delta_{BV}{}^2=0.
\eqno(5.11)
$$
The associated cohomology is the classical Batalin--Vilkovisky cohomology.
Also, by (5.8) 
$$
\delta_{BV} S=0.
\eqno(5.12)
$$

It is interesting to see how the odd symplectic form $\Omega$ behaves 
under a $b$ transform of the $H$ field of the form (2.13).
It turns out that a meaningful comparison of the resulting symplectic form $\Omega'$ 
and the original symplectic form $\Omega$ requires that the superfields $x^a$, $y_a$ 
also must undergo a $b$ transform of the form 
$$
\eqalignno{\vphantom{1\over 2} 
&x'{}^a=x^a, &(5.13a)\cr
\vphantom{1\over 2}
&y'{}_a=y_a+b_{ab}(x)dx^b.&(5.13b)\cr
}
$$
It is then simple to verify that \xxx
$$
\Omega'=\Omega.
\eqno(5.14)
$$ 
If the $2$--form $b$ is closed, then $H'=H$, by (2.4). 
In that case, the $b$ transform is canonical, i. e. 
it leaves the Batalin--Vilkovisky odd symplectic form (5.3) invariant. 

It is similarly interesting to see how the action $S$ behaves under a $b$ transform 
of the $H$ field and of the generalized almost complex structure $\cal J$ of the form 
(2.13), (2.16{\it a--c}). Provided the field redefinition (5.13{\it a, b})
are carried out, one finds that the resulting action $S'$ action equals the 
original one $S$, \xxx
$$
S'=S.
\eqno(5.15)
$$
So, {\it $b$ transform is a duality symmetry of the Hitchin sigma model} \ref{31}.
\footnote{}{}\footnote{${}^5$}{In the untwisted Hitchin model, the $H$ field is 
absent and the action $S$ has no topological term.
In that case the $2$--form $b$ must be closed and $S'=S-2\int_\Sigma x^{(0)*}b$, 
i. e. $S$, $S'$ differ by a topological term. If $b/2\pi$ 
has integer periods and, so, describes a gerbe gauge transformation,
one has $\exp(\sqrt{-1}S')=\exp(\sqrt{-1}S)$ in the quantum path integral.
$b$ transform is then a duality symmetry of the quantum Hitchin sigma model.
See \ref{31}.}

The Batalin--Vilkovisky variation operator $\delta_{BV}$ 
behaves covariantly under $b$ transformation in the following sense. 
Let $H'$ and ${\cal J}'$ be a $b$ transform of the $H$ field and of the generalized 
complex structure $\cal J$ given in (2.13), (2.16{\it a--c}). 
Let $x'{}^a$, $y'{}_a$ be the $b$ transform of the superfields $x^a$, $y_a$ given in 
(5.13{\it a, b}). Let $\delta'_{BV}x'{}^a$, $\delta'_{BV}y'{}_a$ be given by 
(5.10{\it a, b}) with $x^a$, $y_a$ and  $H_{abc}$, $P^{ab}$, $J^a{}_b$
$Q_{ab}$ replaced by $x'{}^a$, $y'{}_a$ $H'{}_{abc}$, $P'{}^{ab}$, $J'{}^a{}_b$
$Q'{}_{ab}$, respectively. Then, one has the relations \xxx
$$
\eqalignno{\vphantom{1\over 2} 
&\delta'_{BV}x'{}^a=\delta_{BV}x^a, &(5.16a)\cr
\vphantom{1\over 2}
&\delta'_{BV}y'{}_a=\delta_{BV}\big(y_a+b_{ab}(x)dx^b\big). &(5.16b)\cr
}
$$
(Compare (5.16{\it a, b}) with (5.13{\it a, b}).) In this sense, 
{\it $b$ transformation is compatible with the Batalin--Vilkovisky $\delta_{BV}$ 
cohomological structure}.

Let see next how one can incorporate branes into the Hitchin sigma model. 
The discussion at the end of sect. 3 
shows that, if we want to make contact with the usual picture of branes, a brane 
contained in the target bulk space should not be seen simply as an ordinary 
submanifold $W$ of the manifold $M$, but rather as a generalized submanifold $(W,F)$ 
of the twisted manifold $(M,H)$. 

A sigma model with branes describes the dynamics of open strings whose ends lie 
on the branes. Therefore, the world sheet $\Sigma$ of our sigma model with branes 
must have a non empty boundary $\partial\Sigma$ and the superfields  
$x$, $y$ must satisfy appropriate boundary conditions. 
These are properly expressed in terms of the pull--back superfields  
$\iota_\partial{}^*x$, $\iota_\partial{}^*y$ (cf. sect. 4). 
Below, for notational simplicity, 
$\iota_\partial{}^*$ will often be tacitly understood.

The superembedding $x$ must be such that $\iota_\partial{}^*x(\Pi T\partial\Sigma)
\subseteq W$. As a consequence, the superfield $x^a$ must satisfy the boundary 
condition \xxx 
$$
x^\rho=0
\eqno(5.17a)
$$
along $\Pi T\partial\Sigma$, in adapted coordinates (cf. sect. 3). 
From a geometrical point of view, it is also 
natural to demand that the pull--back superfields $\iota_\partial{}^*dx$, 
$\iota_\partial{}^*y$ represent a section of the pull-back 
$(\iota_\partial{}^*x)^*\Pi {\cal T}^FW$ of the parity reversed generalized tangent 
bundle $\Pi{\cal T}^FW$ of $W$ (cf. sect. 3). 
On account of (3.6{\it a, b}), one has the further 
boundary condition 
$$
y_i+F_{ij}(x)d_\partial x^j=0
\eqno(5.17b)
$$
along $\Pi T\partial\Sigma$, in adapted coordinates.
(The condition $d_\partial x^\rho=0$ is already implied by (5.17{\it a}).)
The boundary conditions (5.17{\it a, b}) are purely geometrical 
hence {\it kinematical}.

When $M$ contains a brane $W$, the expression (5.3) of the Batalin--Vilkovisky
odd symplectic $\Omega$ is no longer correct.  As it stands, $\Omega$ 
fails to be closed, as in fact
$$
\delta \Omega=-\oint_{\Pi T\partial\Sigma}\mu_\partial\, \hbox{$1\over 2$}
\partial_iF_{jk}(x)\delta x^i\delta x^j\delta x^k.
\eqno(5.18)
$$
The right hand side is a boundary term and can be compensated for by modifying (5.3)
by a boundary term. The resulting expression is
$$
\Omega_{W}=\int_{\Pi T\Sigma}\mu\,\Big[\delta x^a\delta y_a
+\hbox{$1\over 2$}H_{abc}(x)\delta x^a dx^b \delta x^c\Big]
+\oint_{\Pi T\partial\Sigma}\mu_\partial\, 
\hbox{$1\over 2$}F_{ij}(x)\delta x^i\delta x^j.
\eqno(5.19)
$$
Now, $\Omega_{W}$ is a closed functional form,
$\delta\Omega_{W}=0$. In this way, 
one can define antibrackets $(\,,)_W$ in standard fashion. 
The resulting expression is 
$$
\eqalignno{\vphantom{1\over 2}
(F,G)_W&=\int_{\Pi T\Sigma}\mu\,\bigg[
{\delta_r F\over\delta x^a}{\delta_l G\over\delta y_a}-
{\delta_r F\over\delta y_a}{\delta_l G\over\delta x^a}
-H_{abc}(x){\delta_r F\over\delta y_a} 
dx^b {\delta_l G\over\delta y_c}\bigg]&(5.20)\cr 
\vphantom{1\over 2}
&-\oint_{\Pi T\partial\Sigma}\mu_\partial\, F_{ij}(x)
{\delta_r F\over\delta y_i} {\delta_l G\over\delta y_j},&\cr
}
$$
for any two functionals $F$, $G$ of $x^a$, $y_a$.

Now, let us introduce the Hitchin sigma model in the presence of a brane $(W,F)$.
The target space geometry of the bulk $(M,H)$ is specified again by a generalized 
almost complex structure $\cal J$. $(W,F)$ is assumed to be a generalized almost 
complex submanifold of $(M,H,{\cal J})$ (cf. sect. 3). 
This assumption is natural from a geometrical 
point of view. In the representation (2.11), the action of the model reads 
$$
\eqalignno{\vphantom{1\over 2}
S_W&=\int_{\Pi T\Sigma}\mu\,\Big[y_adx^a+\hbox{$1\over 2$}P^{ab}(x)y_ay_b
+\hbox{$1\over 2$}Q_{ab}(x)dx^adx^b+J^a{}_b(x)y_adx^b\Big]&(5.21)\cr
\vphantom{1\over 2}
&-2\int_\Gamma \bar x^{(0)*}H+2\int_\Delta \bar x^{(0)*}F.&\cr
}
$$
Here, $\Gamma$, $\Delta$ are respectively a $3$-- and a $2$--fold such that 
$\Delta\subseteq\partial\Gamma$,
\vfill\eject\noindent 
$$
\partial \Gamma-\Delta=\Sigma, \qquad \partial\Delta=-\partial\Sigma
\eqno(5.22)
$$
$\bar x^{(0)}:\Gamma\rightarrow M$ is an embedding such that 
$\bar x^{(0)}(\Delta)\subseteq W$ and that $\bar x^{(0)}|_\Sigma$
equals the lowest degree $0$ component $x^{(0)}$ of the superembedding $x$.
The last two terms dependent on $H$, $F$ constitute altogether a Wess--Zumino 
like term. Their total value depends on the choice of 
$\bar x^{(0)}$. In the quantum theory, in order to have a 
well defined weight $\exp(\sqrt{-1}S_W)$ in the path integral, 
it is necessary to require that $(H/2\pi,F/2\pi)$ has integer 
relative periods, so that the relative cohomology class $[(H/2\pi,F/2\pi)]$
belongs to the image of $H^3(M,W,\Bbb Z)$ in $H^3(M,W,\Bbb R)$ (cf. sect. 3).
The above definition of the topological term works only if
$(x^{(0)}(\Sigma),x^{(0)}(\partial\Sigma))$ is a relative boundary in $M$. 
If one wants to extend the definition to the general case where 
$(x^{(0)}(\Sigma),x^{(0)}(\partial\Sigma))$
is a relative cycle of $M$, the theory of Cheeger--Simons relative 
differential characters is required \ref{51}. 

The brane action $S_W$ given in (5.21) is the most obvious generalization of the 
no brane action $S$ given in (5.5). It is so designed to yield the same 
expression for the derivatives ${\delta_{l,r} S_W\over\delta x^a}$, 
${\delta_{l,r} S_W\over\delta y_a}$ of $S_W$ and the derivatives 
${\delta_{l,r} S\over\delta x^a}$, ${\delta_{l,r} S\over\delta y_a}$ of $S$
upon imposing a natural boundary condition on the fields as we show next. 
Using the general definition (4.16) to compute ${\delta_{l,r}  S_W\over\delta x^a}$, 
one obtains an integral on $\Pi T\Sigma$ and a boundary integral 
on $\Pi T\partial\Sigma$ of the form 
$$
B.I.=-\oint_{\Pi T\partial\Sigma}\mu_\partial\, 
F_{ij}(x)\big[d_\partial x^j+J^j{}_k(x) d_\partial x^k+P^{j\rho}(x)y_\rho
-P^{jk}F_{kl}(x)d_\partial x^l\big]\delta x^i, 
\eqno(5.23)
$$
in adapted coordinates. The calculation involves repeated applications of the 
relations (3.7{\it a--c}) which follow from $(W,F)$ being a generalized almost 
complex submanifold of $(M,H,{\cal J})$ and the kinematical boundary conditions 
(5.17{\it a, b}). From (5.23), it seems reasonable to demand that 
$$
F_{ij}(x)\big[d_\partial x^j+J^j{}_k(x) d_\partial x^k+P^{j\rho}(x)y_\rho
-P^{jk}F_{kl}(x)d_\partial x^l\big]=0
\eqno(5.24)
$$
along $\Pi T\partial\Sigma$. 
\footnote{}{}\footnote{${}^6$}{Of course, this condition is no longer necessary when 
$F_{ij}=0$. We shall not consider this possibility in the following discussion.}
Unfortunately, this boundary condition
suffers a number of diseases to be discussed below. These are cured 
by replacing (5.24) by a stronger boundary condition implying (5.24), namely
$$
d_\partial x^i+J^i{}_j(x) d_\partial x^j+P^{i\rho}(x)y_\rho
-P^{ij}F_{jk}(x)d_\partial x^k=0
\eqno(5.25)
$$
along $\Pi T\partial\Sigma$. 
\footnote{}{}\footnote{${}^7$}{Exploiting (5.17{\it a, b}), it is straightforward to 
show that both conditions (5.24), (5.25) are adapted covariant.}
Using again the general definition (4.16) 
to compute ${\delta_{l,r} S_W\over\delta y_a}$, one obtains 
again
an integral on $\Pi T\Sigma$ and no boundary integral on $\Pi T\partial\Sigma$. 
Therefore, no further boundary conditions are required.
The boundary conditions (5.24), (5.25) originate from the structure of the brane 
action $S_W$ and are therefore {\it dynamical}.

The expressions of 
${\delta_{l,r} S_W\over\delta x^a}$, ${\delta_{l,r} S_W\over\delta y_a}$
obtained in this way are given respectively by the right hand side of (5.10{\it b}) 
and plus/minus the the right hand side of (5.10{\it a}). 
Using these, 
one can compute the antibrackets $(S_W,S_W)_W$ using (5.20).
The calculation involves repeated applications of the 
relations (3.7{\it a--c}) and the kinematical boundary conditions 
(5.17{\it a, b}) again, and the dynamical boundary condition (5.24) or (5.25).
(One also uses that $d_\partial x^id_\partial x^j=0$ on the $1$--dimensional
manifold $\partial\Sigma$.) Because of (5.24), the boundary term in the 
right hand side of (5.20) vanishes identically. The result is 
$$
\eqalignno{\vphantom{1\over 2} 
(S_W,S_W)_W=
2\int_{\Pi T\Sigma}\mu\,\Big[&-\hbox{$1\over 6$}A_H{}^{abc}(x)y_ay_by_c
+\hbox{$1\over 2$}B_H{}_a{}^{bc}(x)dx^ay_by_c&(5.26)\cr
\vphantom{1\over 2}
&-\hbox{$1\over 2$}C_H{}_{ab}{}^c(x)dx^adx^by_c
+\hbox{$1\over 6$}D_H{}_{abc}(x)dx^adx^bdx^c&\cr
\vphantom{1\over 2}
&-\hbox{$1\over 3$}H_{abc}(x)dx^adx^bdx^c\Big],
&\cr}
$$
where the tensors $A_H$, $B_H$, $C_H$, $D_H$ are given by (2.18{\it a--d}). 
Note that the expression (5.26) of the brackets $(S_W,S_W)_W$ is formally 
identical to the expression (5.7) of the brackets $(S,S)$. 
Hence, $S_W$ satisfies the brane classical Batalin--Vilkovisky master 
equation 
$$
(S_W,S_W)_W=0,
\eqno(5.27)
$$
if (2.17{\it a--d}) hold, i.e. when $\cal J$ is an $H$ twisted generalized complex 
structure. In this way, the brane $(W,F)$ is a generalized complex submanifold of 
the twisted generalized complex manifold $(M,H,{\cal J})$ (cf. sect. 3). 
This shows that {\it the non trivial connection between generalized complex 
geometry and quantization \`a la Batalin--Vilkovisky of the sigma model continues 
to hold also when $M$ contains a brane $W$.} This is so by design. 

The Batalin--Vilkovisky variations $\delta_{BVW}x^a$, $\delta_{BVW}y_a$ 
are given by 
$$
\eqalignno{\vphantom{1\over 2} 
\delta_{BVW} x^a&=(S_W,x^a)_W,&(5.28a)\cr
\vphantom{1\over 2}
\delta_{BVW} y_a&=(S_W,y_a)_W. &(5.28b)\cr
}
$$
Using (5.20), we can then derive the expressions of the brane 
Batalin--Vilkovisky variations $\delta_{BVW} x^a$, $\delta_{BVW} y_a$.
They are given by the same formulae as those of the no brane 
Batalin--Vilkovisky variations $\delta_{BV} x^a$, $\delta_{BV} y_a$
given in (5.10{\it a, b}). Again, this is so by design.
As $\delta_{BV}$ (cf. eq. (5.11)), the brane Batalin--Vilkovisky variation operator
$\delta_{BVW}$ is nilpotent 
$$
\delta_{BVW}{}^2=0.
\eqno(5.29)
$$
The associated cohomology is the brane classical Batalin--Vilkovisky cohomology.
This relation is not the trivial restatement of (5.11), which it may appear at 
first glance, since, in the presence of branes,  the superfields
$x^a$, $y_a$ obey the boundary conditions derived above. 
(The natural question about the compatibility of the boundary conditions 
with the Batalin--Vilkovisky cohomological structure will be discussed in 
the next section.) Also, one has \xxx 
$$
\delta_{BVW} S_W=0,
\eqno(5.30)
$$
by (5.27).

It is interesting to check whether the incorporation of branes into 
the Hitchin sigma model is compatible with the $b$ transformation symmetry. 

To begin with, let us see how the odd symplectic form $\Omega_W$ behaves 
under a $b$ transform of the $H$ and $F$ fields of the form (2.13), (3.5).
Again, a meaningful comparison of the resulting symplectic form $\Omega'_W$ 
and the original symplectic form $\Omega_W$ requires that the superfields 
$x^a$, $y_a$ undergo the  $b$ transform (5.13{\it a, b}). 
From (5.19), it is straightforward to verify that \xxx
$$
\Omega'_W=\Omega_W,
\eqno(5.31)
$$ 
generalizing (5.14).

Next, let us see how the action $S_W$ behaves under a $b$ transform of the $H$ 
and $F$ fields and of the generalized almost complex structure $\cal J$ of the form 
(2.13), (3.5), (2.16{\it a--c}). Using (5.13{\it a, b}), one finds that \xxx
$$
S'_W=S_W, 
\eqno(5.32)
$$
generalizing (5.15). So, {\it $b$ transform is a duality symmetry of the Hitchin 
sigma model even in the presence of a brane.} 

Expectedly, the brane Batalin--Vilkovisky variation operator $\delta_{BVW}$ 
behaves covariantly under $b$ transformation as in the no brane case. 
Relations (5.16{\it a, b}) hold true with $\delta_{BV}$, $\delta'_{BV}$
replaced by $\delta_{BVW}$, $\delta'_{BVW}$, respectively, where $\delta'_{BVW}$ 
is defined analogously to $\delta'_{BV}$. Thus, 
{\it $b$ transformation is compatible with the Batalin--Vilkovisky cohomological 
structure also in the brane case}.

In the presence of a brane $W$, the superfields $x^a$, $y_a$ obey the boundary 
conditions (5.17{\it a, b}) and (5.24) or (5.25). It is important to check 
whether these conditions are invariant under $b$ transformation. 

Let us consider first the kinematical boundary conditions (5.17{\it a, b}). 
Requiring their $b$ invariance amounts to demanding that
whenever the $x^a$, $y_a$ satisfy (5.17{\it a, b}) with respect to the 
brane field $F$, their $b$ transforms $x'{}^a$, $y'{}_a$,
given by (5.13{\it a, b}), satisfy (5.17{\it a, b}) with respect to the 
$b$ transformed brane field $F'$, given by (3.5). It is trivial to check that 
this is indeed the case. 

Let us consider next the dynamical boundary conditions (5.24) or (5.25). 
Again, requiring their $b$ invariance amounts to demanding that
whenever the $x^a$, $y_a$ satisfy (5.24) or (5.25) with respect to the 
fields $H$ and $F$ and the generalized almost complex structure $\cal J$, 
their $b$ transforms $x'{}^a$, $y'{}_a$,
given by (5.13{\it a, b}), satisfy (5.24) or (5.25) with respect to the 
$b$ transformed fields $H'$ and $F'$ and the $b$ transformed 
generalized almost complex structure $\cal J'$, given by (2.13), (3.5),
(2.16{\it a--c}). It is trivial to check that this is the case
for the boundary condition (5.25) but not for the boundary condition (5.24). 
The latter therefore breaks $b$ invariance. $b$ transformation symmetry 
is one of the most basic features of both generalized complex geometry and the
Hitchin sigma model. Renouncing to it seems to be out of question. 
This indicates that {\it (5.25) is the appropriate dynamical boundary condition}.

\titlebf{6. The classical BV cohomology in the presence of branes}

\par
One of the most interesting aspects of the Hitchin sigma model is the 
associated classical Batalin--Vilkovisky cohomology, which describes 
the observables of the model at semiclassical level \ref{36,37}. 
As we have seen above, when a brane $W$ is present, 
the expressions of the brane Batalin--Vilkovisky variations $\delta_{BVW} x^a$, 
$\delta_{BVW} y_a$ are the same as those of the no brane variations 
$\delta_{BV} x^a$, $\delta_{BV} y_a$ given in (5.10{\it a, b}). One would thus 
conclude that the no brane and brane classical Batalin--Vilkovisky cohomology are 
identical. However, one should keep in mind that, in the brane case the basis fields 
$x^a$, $y_a$ of the model obey the kinematical boundary conditions (5.17{\it a, b})
and the dynamical boundary conditions (5.24) or (5.25). It is therefore important to 
check the compatibility of the boundary conditions and the overall 
Batalin--Vilkovisky cohomological structure. 

Concretely, this is done as follows. 
The boundary conditions have the general form 
$$
{\eul F}(\iota_\partial{}^*x, \iota_\partial{}^* y)=0,
\eqno(6.1)
$$
where $\eul F$ is some functional of the boundary superfields. 
The boundary conditions are compatible with 
Batalin--Vilkovisky cohomological structure provided 
$$
\delta_{BVW} {\eul F}(\iota_\partial{}^* x, \iota_\partial {}^*y)=0,
\eqno(6.2)
$$
when (6.1) holds, where the left hand side is computed using 
(5.10{\it a, b}) and the fact that $\delta_{BVW} \iota_\partial{}^*=
\iota_\partial{}^*\delta_{BVW}$. If (6.2) does not hold,
then (6.2) becomes a new set of boundary conditions to be added to the original 
ones. The compatibility check then must be carried out again 
from the beginning with the
enlarged set of boundary conditions so obtained. The process must be continued 
until (6.2) is satisfied identically. Alternatively, one 
may replace the boundary conditions (6.1) with stronger boundary 
conditions implying (6.1), for which (6.2) holds identically.
(These will be expressed in terms of some functional 
$\hat {\eul F}$ different from $\eul F$.)

We consider first the kinematical boundary conditions (5.17{\it a, b}). 
Their compatibility with the Batalin--Vilkovisky cohomological structure
requires that 
$$
\eqalignno{\vphantom{1\over 2}    
&\delta_{BVW} x^\rho=0,&(6.3a)\cr
\vphantom{1\over 2}
&\delta_{BVW}y_i-F_{ij}(x)d_\partial \delta_{BVW}x^j
+\partial_kF_{ij}(x)\delta_{BVW}x^kd_\partial x^j=0&(6.3b)\cr
}
$$
along $\Pi T\partial\Sigma$, in adapted coordinates, when
(5.17{\it a, b}) hold. Using (3.7{\it a--c}) and (5.17{\it a, b}) 
it is straightforward to verify that these relations do indeed hold true.

We consider next the dynamical boundary conditions (5.24) or (5.25). In the 
previous section, we saw that the basic requirement of $b$ transformation symmetry 
invariance of the dynamical boundary conditions rules out (5.24) and selects 
(5.25) as the only consistent condition. We now shall show that the requirement 
of compatibility of the dynamical boundary conditions with the Batalin--Vilkovisky 
cohomological structure leads to the very same conclusion.

By (5.10{\it a}), the condition (5.24) can be written concisely as 
$$
F_{ij}(x)\delta_{BVW}x^j=0
\eqno(6.4)
$$
along $\Pi T\partial\Sigma$.  Proceeding as above, its compatibility 
with the Batalin--Vilkovisky cohomological structure
requires that \xxx
$$
\partial_k F_{ij}(x)\delta_{BVW}x^k\delta_{BVW}x^j=0
\eqno(6.5)
$$
along $\Pi T\partial\Sigma$, when (5.17{\it a, b}) and (5.24) hold. 
Explicitly written, (6.5) reads
$$
\big(\partial_kF_{ij}+\partial_jF_{ki}\big)P^{k\sigma}(x)y_\sigma
\big[d_\partial x^j+J^j{}_l(x) d_\partial x^l+
\hbox{$1\over 2$}P^{j\rho}(x)y_\rho
-P^{jl}F_{lm}(x)d_\partial x^m\big]=0
\eqno(6.6)
$$
along $\Pi T\partial\Sigma$. (6.6) is not satisfied identically, when 
(5.17{\it a, b}) and (5.24) hold. According to the procedure described above,  
(6.6) should be added as a new boundary condition. 
From (5.29), (6.5), it is easy to see that no further boundary conditions 
would be required by compatibility. 
The alternative condition (5.25) can be written 
concisely 
$$
\delta_{BVW}x^j=0
\eqno(6.7)
$$
along $\Pi T\partial\Sigma$. By (5.29), (6.7), the boundary condition (5.25) 
satisfies the compatibility requirements (6.2) trivially.

In conclusion, (5.24) involves the introduction of the new dynamical boundary 
condition (6.6). (6.6) is rather complicated and does not seem to have any 
obvious natural interpretation. (5.25), conversely, does not have this undesirable 
feature. This analysis provides conclusive evidence that (5.25), not (5.24), 
is the appropriate consistent dynamical boundary condition of Hitchin 
sigma model. 
Combining (6.3{\it a}), (6.7) one gets the boundary condition \xxx
$$
\delta_{BVW}x^a=0
\eqno(6.8)
$$
along $\Pi T\partial\Sigma$, where the left hand side is given by (5.10{\it a}). 
This seems to be a more transparent set of boundary conditions, also in the light  
of the fact that the $y$ classical field equations of the Hitchin sigma model can 
be written precisely as $\delta_{BVW}x^a=0$ on $\Pi T\Sigma$.

In ref. \ref{31}, the classical Batalin--Vilkovisky cohomology of the no brane  
Hitchin sigma model was studied. In particular, it was shown that there exists a 
natural homomorphism of the generalized Dolbeault cohomology of target space
into the Batalin--Vilkovisky cohomology depending on a choice of a singular 
supercycle $Z$ of the world sheet $\Sigma$. 
The analysis was based on an adaptation of the well--known descent formalism
of gauge theory. The homomorphism was constructed by associating 
with a representative of a generalized Dolbeault cohomology class 
a local superfield and integrating the latter on $Z$ and by showing that 
the results was a representative of a Batalin--Vilkovisky cohomology
class. It was also found that the generalized Dolbeault cohomology contains the 
cohomology of the generalized deformation complex as a subcohomology.
We shall not repeat the details of this analysis. The interested
reader is invited to read sect. 6 of ref. \ref{31} for a thorough discussion 
of these matters. 

One may wonder whether and how the presence of branes modifies the classical 
Batalin--Vilkovisky cohomology. One may think, for instance, that a relative 
version of the generalized Dolbeault cohomology may be the appropriate 
cohomology in this case, since relative cohomology involves typically a space 
and a subspace, here the target bulk and the brane. This is unlikely to work out 
for the following simple reason. A generalized complex submanifold of a twisted 
generalized complex manifold is {\it not} itself a twisted generalized complex 
manifold in general, while any reasonable relative version of a cohomology requires 
that the two spaces involved carry the same type of structure. 

A more likely scenario is the following. In the presence of branes, the  
Batalin--Vilkovisky cohomology will contain a sector reproducing the no
brane cohomology associated with the generalized Dolbeault cohomology of the bulk 
and a further sector based on some cohomological structure of the brane, whose 
nature is to be determined. We devote the rest of this section to the 
analysis of this matter. The discussion parallels 
that of the no brane cohomology of ref. \ref{31} with some 
significant differences to be discussed below. 

The construction expounded in the following involves the singular chain complex of 
$\partial\Sigma$. Since we deal with boundary de Rham superfields throughout
(cf. sect. 4), it is convenient to use the boundary singular superchains formalism.
A boundary singular superchain $C$ is a doublet formed by a $0$--, 
$1$--dimensional singular chain $C_{(0)}$, $C_{(1)}$ of $\partial \Sigma$ 
organized as a formal chain sum $C=C_{(0)}+C_{(1)}$. The nilpotent singular boundary 
operator $\partial$ of $\partial \Sigma$ extends to boundary superchains in 
obvious fashion by setting $(\partial_\partial C)_{(0)}=\partial C_{(1)}$, 
$(\partial_\partial C)_{(1)}=0$. A boundary singular supercycle $Z$ is a superchain 
such that $\partial_\partial Z=0$.
A boundary de Rham superfield $\chi$ can be integrated on a boundary superchain $C$:
$$
\int_C\chi=\int_{C_{(0)}}\chi^{(0)}
+\int_{C_{(1)}}ds\chi^{(1)}{}_s(s).
\eqno(6.9)
$$
Stokes' theorem holds, $\int_C d_\partial \chi=\int_{\partial_\partial C} \chi$.
In particular, one has $\int_Zd_\partial \chi=0$ for any boundary supercycle $Z$.

We call a boundary de Rham superfield $X$ local, if it is a local functional 
of the pull--back of the basic superfields $\iota_\partial{}^*x$, 
$\iota_\partial{}^*y$. Let $X$ be some local boundary 
superfield and let there be another local boundary superfield $Y$ such that 
$$
\delta_{BVW}X=d_\partial Y.
\eqno(6.10)
$$
Thus, $X$ defines a mod $d_\partial$ Batalin--Vilkovisky cohomology class.
Then, if $Z$ is a boundary singular supercycle, one has \xxx
$$
\delta_{BVW}\int_ZX=\int_Zd_\partial Y=0,
\eqno(6.11)
$$
by Stokes' theorem. It follows that \xxx
$$
\langle Z, X\rangle_\partial =\int_ZX
\eqno(6.12)
$$
defines a Batalin--Vilkovisky cohomology class. A standard analysis shows that this 
class depends only on the mod $d_\partial $ Batalin--Vilkovisky cohomology 
class of $X$. 
So, one may obtain Batalin--Vilkovisky cohomology classes by constructing 
boundary local superfields $X$ satisfying (6.10).

Define \xxx
$$
\partial_\partial=\hbox{$1\over 2$}\big[d_\partial
-\sqrt{-1}(\delta_{BVW}-d_\partial)\big]
\eqno(6.13)
$$
and its complex conjugate $\overline\partial_\partial$, where here and below 
the operator $\delta_{BVW}$ is tacitly restricted to the space of local boundary 
superfields. From (5.29), using that $d_\partial\delta_{BVW}
+\delta_{BVW}d_\partial=0$, it is immediate to check that \xxx
$$
\eqalignno{
\vphantom{1\over 2}
\partial_\partial{}^2&=0,&(6.14a)\cr
\vphantom{1\over 2}
\overline\partial_\partial{}^2&=0,&(6.14b)\cr
\vphantom{1\over 2}
\partial_\partial\overline\partial_\partial
+\overline\partial_\partial\partial_\partial&=0.&(6.14c)\cr
}
$$
From (6.5), one has further \xxx
$$
\eqalignno{
\vphantom{1\over 2}
d_\partial&=\partial_\partial+\overline\partial_\partial,&(6.15a)\cr
\vphantom{1\over 2}
\delta_{BVW}&=\partial_\partial+\overline\partial_\partial
+\sqrt{-1}(\partial_\partial-\overline\partial_\partial).
&(6.15b)\cr
}
$$

The operator $\overline\partial_\partial$ acts on the space of local boundary 
superfields, carries degree $1$, by (6.13), and it squares to $0$, by (6.14{\it b}). 
Therefore, one can define a $\overline\partial_\partial$
local boundary superfield cohomology in obvious fashion.

Let $X$ be a local boundary superfield such that 
$$
\overline\partial_\partial X=0.
\eqno(6.16)
$$
$X$ defines a $\overline\partial_\partial$ local boundary superfield cohomology class. 
By (6.15{\it a, b}), $X$ satisfies (6.10) with $Y=(1+\sqrt{-1})X$. So, as shown above, 
for any boundary supercycle $Z$, $\langle Z, X\rangle_\partial$ 
defines a Batalin--Vilkovisky cohomology class. If $X=\overline\partial_\partial U$ 
for some local boundary superfield $U$, so that the corresponding 
$\overline\partial_\partial$ cohomology class is trivial, then 
$\langle Z, X\rangle_\partial={1\over 2}\sqrt{-1}\delta_{BVW}
\langle Z, U\rangle_\partial$, by (6.13) and Stokes' theorem and, so, 
the corresponding Batalin--Vilkovisky class 
is trivial too. Therefore, for any boundary singular supercycle $Z$,
there is a well-defined homomorphism from the $\overline\partial_\partial$ 
local boundary superfield cohomology into the Batalin--Vilkovisky cohomology. This 
homomorphism depends only on the boundary singular homology class of $Z$.

Adapting the analysis of ref. \ref{31}, 
we carry out the construction of the brane sector of the 
Batalin--Vilkovisky cohomology by associating with a given field of the brane 
a local boundary superfield and imposing that the latter satisfies (6.16). 
Taking into account the boundary conditions
(5.17{\it a, b}), we see that the basic boundary superfields at our disposal 
are the $x^i$, $y_\rho$, in adapted coordinates. 
(Here and below, the pull--back operator $\iota_\partial{}^*$
is tacitly understood.) 
Tentatively, we may consider local boundary superfields of the form
$$
X_\Upsilon=\sum_{p,q\geq 0}\hbox{$1\over p!q!$}
\Upsilon^{\rho_1\dots\rho_p}{}_{i_1\dots i_q}(x)
y_{\rho_1}\ldots y_{\rho_p}d_\partial x^{i_1}\ldots d_\partial x^{i_q},
\eqno(6.17)
$$
where $\Upsilon\in\oplus_{p,q\geq 0}C^\infty(\wedge^p NW\otimes
\wedge^q T^*W\otimes\Bbb C)$, $NW=TM|_W/TW$ being the normal bundle of $W$.
This is however too naive. In fact, under a change of 
adapted coordinates (cf. sect. 3), the $d_\partial x^i$ transform into themselves,
since $d_\partial x^\rho=0$ by (5.17{\it a}), while the $y_\rho$ do not 
but mix with the $F_{ij}(x)d_\partial x^j$, since $y_i=-F_{ij}(x)d_\partial x^j$ 
by (5.17{\it b}). For this reason, an object of the form (6.17) would not be 
adapted covariant (cf. sect. 3). To get adapted covariant expressions, 
the components $\Upsilon^{\rho_1\dots\rho_p}{}_{i_1\dots i_q}$ of the 
brane field $\Upsilon$ must have more exotic transformation properties than 
those simplemindedly implied above. Formally, these are given as follows.
$\Upsilon\in{\scri X}_{W0}{}^*=\oplus_{n\geq 0}{\scri X}_{W0}{}^n$,  
where ${\scri X}_{W0}{}^n$ is the space of the formal adapted covariant 
objects 
$$
\Upsilon=\sum_{p,q\geq 0,p+q=n}\hbox{$1\over p!q!$}
\Upsilon^{\rho_1\dots\rho_p}{}_{i_1\dots i_q}
\eta_{\rho_1}\ldots \eta_{\rho_p}\xi^{i_1}\ldots \xi^{i_q}
\eqno(6.18)
$$
$\xi^i$, $\eta_\rho$ being anticommuting objects transforming as
$$
\eqalignno{
\vphantom{1\over 2}\xi'^i&=\partial_j t'^i\xi^j, &(6.19a)\cr
\vphantom{1\over 2}\eta'{}_\rho&=\partial'{}_\rho t^\sigma\eta_\sigma
- \partial'{}_\rho t^iF_{ij}\xi^j, &(6.19b)\cr
}
$$
under a change of adapted coordinates. The resulting transformation properties
of the component fields $\Upsilon^{\rho_1\dots\rho_p}{}_{i_1\dots i_q}$ are {\it not }
tensorial and in explicit form are rather messy. 

We note that all terms in the right hand side of (6.17) with $q\geq 1$ vanish,
since $d_\partial x^id_\partial x^j=0$ on the $1$--dimensional manifold 
$\partial\Sigma$. This property will be used repeatedly
to suitably reshape the expressions derived below.

A simple calculation using (5.10{\it a, b}), (3.7{\it a--c}), (5.17{\it a, b})
yields
$$
\eqalignno{
\vphantom{1\over 2} 
\delta_{BVW}x^i&=
d_\partial x^i+P^{i\rho}(x)y_\rho
+\big(J^i{}_j-P^{ik}F_{kj}\big)(x)d_\partial x^j,&(6.20a)\cr
\vphantom{1\over 2}
\delta_{BVW} y_\rho&=d_\partial y_\rho
+\hbox{$1\over 2$}\partial_\rho P^{\sigma\tau}(x)y_\sigma y_\tau
+\big(\partial_\rho J^\sigma{}_i
-\partial_iJ^\sigma{}_\rho &(6.20b)\cr
\vphantom{1\over 2}
&+P^{\sigma j}H_{ij\rho}-F_{ij}\partial_\rho P^{j\sigma}\big)(x)y_\sigma dx^i
+J^\sigma{}_\rho(x)d_\partial y_\sigma.
&\cr
}
$$
From these relations and (6.13), (6.17), one finds
$$
\eqalignno{
\vphantom{1\over 2}
\overline\partial_\partial X_\Upsilon
&=\sum_{p,q\geq 0}\hbox{$1\over p!q!$}
\overline\partial_W \Upsilon^{\rho_1\ldots \rho_p}{}_{i_1\ldots i_q}(x)
y_{\rho_1}\cdots y_{\rho_p}d_\partial x^{i_1}\cdots d_\partial x^{i_q}&(6.21)\cr
\vphantom{1\over 2}
&+\sum_{p,q\geq 0}\hbox{$1\over p!q!$}
K_W{}^\sigma \Upsilon^{\rho_1\ldots \rho_p}{}_{i_1\ldots i_q}(x)
d_\partial y_\sigma y_{\rho_1}\cdots y_{\rho_p}
d_\partial x^{i_1}\cdots d_\partial x^{i_q}, &\cr
}
$$
where
$$
\eqalignno{
\vphantom{1\over 2}
\overline\partial_W \Upsilon^{\rho_1\ldots \rho_p}{}_{i_1\ldots i_q}=
\hbox{$1\over 2$}\Big\{&(-1)^p q\Big[\partial_{[i_1}
\Upsilon^{\rho_1\ldots \rho_p}{}_{i_2\ldots i_q]}
&(6.22a)\cr
\vphantom{1\over 2}
&+\sqrt{-1}\big(J^j{}_{[i_1}-P^{jk}F_{k[i_1}\big)\partial_{|j|}
\Upsilon^{\rho_1\ldots \rho_p}{}_{i_2\ldots i_q]}\Big]
&\cr
\vphantom{1\over 2}
&+\hbox{$1\over 2 $}(-1)^pq(q-1)\sqrt{-1}
M_{[i_1}{}^j{}_{i_2}\Upsilon^{\rho_1\ldots \rho_p}{}_{|j|i_3\ldots i_q]}
&\cr
\vphantom{1\over 2}
&+(-1)^ppq\sqrt{-1}M_{[i_1}{}^{[\rho_1}{}_{|\sigma|}
\Upsilon^{|\sigma|\rho_2\ldots \rho_p]}{}_{i_2\ldots i_q]}&\cr
\vphantom{1\over 2}
&-p\sqrt{-1}\Big[P^{[\rho_1|j|}\partial_j 
\Upsilon^{\rho_2\ldots \rho_p]}{}_{i_1\ldots i_q}
&\cr
\vphantom{1\over 2}
&-\hbox{$1\over 2 $}(p-1)\partial_\sigma P^{[\rho_1\rho_2}
{\Upsilon}^{|\sigma|\rho_3\ldots \rho_p]}{}_{i_1\ldots i_q}
&\cr
\vphantom{1\over 2}
&-q\partial_{[i_1}P^{j[\rho_1}
{\Upsilon}^{\rho_2\ldots \rho_p]}{}_{|j|i_2\ldots i_q]}\Big]
&\cr
}
$$
$$
\eqalignno{
\vphantom{1\over 2}
&+\hbox{$1\over 2 $}q(q-1)\sqrt{-1}Z_{[i_1i_2|\sigma|}
\Upsilon^{\sigma\rho_1\ldots \rho_p}{}_{i_3\ldots i_q]}\Big\},&\cr
}
$$
with
$$
\eqalignno{
\vphantom{1\over 2}
M_i{}^k{}_j&= -\partial_i\big(J^k{}_j-P^{kl}F_{lj}\big)+
\partial_j\big(J^k{}_i-P^{kl}F_{li}\big),&(6.22b)\cr
\vphantom{1\over 2}
M_i{}^\sigma{}_\rho&=\partial_iJ^\sigma{}_\rho
-\partial_\rho J^\sigma{}_i
-P^{\sigma j}H_{ij\rho}+F_{ij}\partial_\rho P^{j\sigma},&(6.22c)\cr
\vphantom{1\over 2}
Z_{ij\rho}&=\partial_i Q_{j\rho}+\partial_jQ_{\rho i}
+\partial_\rho Q_{ij}+\partial_i\big(F_{jk}J^k{}_\rho\big)-
\partial_j\big(F_{ik}J^k{}_\rho\big)
&(6.22d)\cr
\vphantom{1\over 2}
&+F_{ik}\partial_\rho J^k{}_j-F_{jk}\partial_\rho J^k{}_i
+F_{ik}F_{jl}\partial_\rho P^{kl}
&\cr
\vphantom{1\over 2}
&-\big(J^k{}_i-P^{kl}F_{li}\big)H_{jk\rho}
+\big(J^k{}_j-P^{kl}F_{lj}\big)H_{ik\rho},
&\cr
}
$$
the brackets $[\cdots]$ denoting full antisymmetrization of all enclosed 
indices except for those between bars $|\cdots|$, and 
$$
\eqalignno{
\vphantom{1\over 2}
K_W{}^\sigma\Upsilon^{\rho_1\ldots \rho_p}{}_{i_1\ldots i_q}=
\hbox{$1\over 2$}\Big\{&\big(\delta^\sigma{}_\tau+\sqrt{-1}J^\sigma{}_\tau\big)
{\Upsilon}^{\tau\rho_1\ldots \rho_p}{}_{i_1\ldots i_q}
&(6.22e)\cr
\vphantom{1\over 2}
&+(-1)^p\sqrt{-1}P^{\sigma j}
{\Upsilon}^{\rho_1\ldots \rho_p}{}_{ji_1\ldots i_q}\Big\}.
&\cr
}
$$
So, $X_\Upsilon$ satisfies (6.16), if \xxx
$$
\eqalignno{
\vphantom{1\over 2}
\overline\partial_W \Upsilon^{\rho_1\ldots \rho_p}{}_{i_1\ldots i_q}&=0,&(6.23a)\cr
\vphantom{1\over 2}
K_W{}^\sigma \Upsilon^{\rho_1\ldots \rho_p}{}_{i_1\ldots i_q}&=0.&(6.23b)\cr
}
$$
It can be verified that eqs. (6.23{\it a, b}) are jointly adapted covariant.  
Note that they are so only together, since in fact 
eq. (6.23{\it a}) is not by itself adapted covariant if 
eq. (6.23{\it b}) is not taken into account. 

It is straightforward to show that, for fixed $\rho$, eq. (6.22{\it e}) defines a 
linear operator $K_W{}^\rho:{\scri X}_{W0}{}^n\rightarrow 
{\scri X}_{W0}{}^{n-1}$, for $n\geq 0$. 
Let ${\scri X}_W{}^n$ be the intersection of the kernels of the operators 
$K_W{}^\rho$ with $\rho=1,~\ldots, d-\dim W$, for $n\geq 0$.
Let further ${\scri X}_W{}^*=\oplus_{n\geq 0}{\scri X}_W{}^n$. 
Eq. (6.23{\it b}) then states that $\Upsilon\in{\scri X}_W{}^*$.

Using (6.22{\it a--e}), by a very lengthy algebraic verification
exploiting systematically (2.17{\it a--d}) and (3.7{\it a--c}), one 
finds that eqs. (6.22{\it a--d}) define a linear operator 
$\overline\partial_W:{\scri X}_W{}^n \rightarrow {\scri X}_W{}^{n+1}$
and that \xxx
$$
\overline\partial_W{}^2=0 \qquad \hbox{on ${\scri X}_W{}^*$}.
\eqno(6.24)
$$
Thus, the pair $({\scri X}_W{}^*,\overline\partial_W)$ is a cochain 
complex, with which there is associated a cohomology 
$H^*({\scri X}_W{}^*,\overline\partial_W)$, which we call 
the generalized complex submanifold cohomology of $W$. 
Eq. (6.23{\it a}) then states that $\Upsilon\in{\scri X}_W{}^*$
is a representative of a class of this cohomology.

It is easy to see that {\it eq. (6.17) defines a homomorphism of the 
generalized complex submanifold cohomology of the brane $W$
into the $\overline\partial_\partial$ local boundary superfield 
cohomology}. Recall that the latter is embedded in a sector of the classical 
Batalin--Vilkovisky cohomology associated with the presence of branes. 
Therefore, 
{\it the brane sector of 
\it the classical Batalin--Vilkovisky cohomology is related non trivially to
\it the generalized complex submanifold cohomology of $W$.}

${\scri X}_W{}^*$ is actually a graded algebra and $\overline\partial_W$
is a derivation on this algebra. As a consequence, 
$H^*({\scri X}_W{}^*,\overline\partial_W)$ has a canonical ring structure.
The cohomology ring $H^*({\scri X}_W{}^*,\overline\partial_W)$ characterizes the 
generalized complex submanifold $W$. To the best of our knowledge, 
it was hitherto unknown. 

The investigation of the intrinsic content of 
$H^*({\scri X}_W{}^*,\overline\partial_W)$, obscured by the adapted coordinate
expressions in terms of which we defined it, is beyond the scope of this paper
and should better be left to the mathematicians. 
The remarkable connections emerging here between the brane sector of the 
Batalin--Vilkovisky cohomology of the Hitchin sigma model on one hand and 
various aspects of Hitchin's generalized complex geometry and Gualtieri's
notion of generalized complex submanifold on the other should however be emphasized
for its field theoretic and physical mathematical interest.

\titlebf{7. Discussion}

\par

The Alexandrov--Kontsevich--Schwartz--Zaboronsky formalism of ref. \ref{52} 
is a method of constructing solutions of the Batalin--Vilkovisky classical 
master equation directly, without starting from a classical action with a 
set of symmetries, as is done in the Batalin--Vilkovisky framework. 
In ref. \ref{38}, using such 
formalism, Cattaneo and Felder managed to obtain the Batalin--Vilkovisky action 
of the Poisson sigma model. In ref. \ref{39}, the same authors
extended their analysis by including branes in the target Poisson manifold and 
showed that branes had to be coisotropic submanifolds. (See also refs. \ref{53,54}
for a related study.) 
In spirit, their approach is very close to the one of the present paper. 
This calls for a comparison of their results and those of we have obtained. 

We consider first the case where branes are absent
and discuss the incorporation of branes later.
Following \ref{38}, we view the standard Poisson sigma model as a field theory
whose base space, target space and field configuration space are respectively 
a two dimensional surface $\Sigma$, a Poisson manifold $M$ with Poisson bivector $P$ 
and the space of bundle maps $\phi=(x,y):\Pi T\Sigma\to\Pi T^*M$. 

The supermanifold $\Pi T^*M$ has a canonical odd symplectic structure, or 
$P$--structure, defined by the canonical odd symplectic form $\omega
=du_adt^a$. With $\omega$, there are associated canonical odd Poisson 
brackets $(\,,)_\omega$ in standard fashion.
The algebra of functions on $\Pi T^*M$ with the odd brackets $(\,,)_\omega$
is isomorphic to the algebra of multivector 
fields on $M$ with the standard Schoutens--Nijenhius brackets. 

The space of field configurations inherits an odd symplectic structure from that 
of $\Pi T^*M$ and, so, it also carries a $P$--structure. 
The associated odd symplectic form $\Omega$ is obtained from $\omega$ 
by integration over $\Pi T\Sigma$ with respect to the usual supermeasure $\mu$ 
(cf. eq. (4.5)) and is given by the first term in the right hand side of (5.3).
With $\Omega$, there are associated odd Poisson brackets $(\,,)_\Omega$
over the algebra of functions on field configuration space, called 
Batalin--Vilkovisky antibrackets in the physical literature.

The Poisson bivector field $P$ of $M$ can be identified 
with a function on $\Pi T^*M$ satisfying $(P,P)_\omega=0$. 
Its Hamiltonian vector field is an odd vector field $Q_P$ 
such that $Q_P{}^2=0$ and defines a so called $Q$--structure on $\Pi T^*M$. 

The Poisson bivector $P$ yields a function $S'$ on 
field configuration space, again by integration over 
$\Pi T\Sigma$ with respect to $\mu$, satisfying $(S',S')_\Omega=0$.
$S'$ is given by the second term in the right hand side of (5.5). 
Its Hamiltonian vector field is an odd vector field $Q_{S'}$ such that 
$Q_{S'}{}^2=0$, and, so, it defines a $Q$--structure on field configuration space.

The base space $\Pi T\Sigma$ carries also a canonical $Q$--structure
$d$ corresponding to the usual de Rham differential on $\Sigma$. $d$ induces 
a $Q$--structure $d$ on field space in obvious fashion. $d$ is Hamiltonian, 
as indeed $d=Q_{S_0}$, where $S_0$ is the function on field configuration space
defined by the first term in the right hand side of (5.5). $S_0$ satisfies 
$(S_0,S_0)_\Omega=0$.

One verifies that $(S',S_0)_\Omega=0$. The sum $S=S_0+S'$ thus satisfies 
$(S,S)_\Omega=0$. $S$ is nothing but the Batalin--Vilkovisky 
action of the Poisson sigma model satisfying the Batalin--Vilkovisky master
equation. Its Hamiltonian vector field $\delta_{BV}=Q_S$ is the 
Batalin--Vilkovisky variation operator. 

In the untwisted case $H=0$, 
the construction of the Hitchin sigma model of ref. \ref{31} closely parallels 
that of the Poisson sigma model outlined above. The field space is the same as that
of the Poisson sigma model. The field space odd symplectic form $\Omega$
is also the same, 
being derived from the same 
odd symplectic form $\omega$ of $\Pi T^*M$ exactly in the same way. The relevant 
difference is that the role of the Poisson bivector $P$ is played here by the 
generalized complex structure $\cal J$, which does not correspond to any multivector 
in general, but which still corresponds to some structure of target space. The overall 
framework is otherwise totally analogous. 

In the twisted case $H\not=0$, the situations is more 
complicated.  If $H=d_MB$ is exact, we can $b$ transform $H$ to $0$ by setting 
$b=B$ (cf. eq. (2.4)), recovering in this way the untwisted case, and follow the 
lines outlined above. If $H$ is not exact, this is no longer possible.  The symplectic 
form $\Omega$ given by (5.5) cannot be straightforwardly obtained from a 
corresponding structure of $\Pi T^*M$. Further, the action functionals contains 
a topological Wess--Zumino term and so it involves an extra geometrical datum, 
the extension of the embedding field $x$ to a 3--dimensional manifold $\Gamma$ 
bounded by $\Sigma$. The spirit of the Alexandrov--Kontsevich--Schwartz--Zaboronsky 
formalism, however, is still fully discernible, in spite of the technical 
differences. 

Let us now examine the case when branes are present. In ref. \ref{39}, the Poisson 
sigma model in the presence of branes is studied. It is found that branes are 
coisotropic submanifolds $W$ of the target Poisson manifold $(M,P)$. 
For a given brane $W$, the boundary conditions impose that the world sheet field
$\phi$ maps $\Pi T\partial\Sigma$ into the parity reversed conormal bundle 
$\Pi N^*W$. Writting $\phi=(x,y)$ as usual and using adapted coordinates, the 
conditions are given by (5.17{\it a, b}) with $F_{ij}=0$. 

In this work, we study similarly the Hitchin sigma model in the presence of branes.
We find that branes are generalized complex submanifolds $(W,F)$ of the target 
twisted generalized complex manifold $(M,H,{\cal J})$ (cf. sect. 3).
For a given $(W,F)$, the boundary conditions impose that the world sheet fields $\phi$ 
maps $\Pi T\partial\Sigma$ into the the parity reversed generalized tangent 
bundle $\Pi{\cal T}^FW$ of $W$ (cf. sects. 3, 5). 
Writting $\phi=(x,y)$ and using adapted coordinates again, the boundary conditions 
are given by (5.17{\it a, b}) and (5.25). 

Now, we recall that not all Poisson manifolds $(M,P)$ are particular cases of 
twisted generalized complex manifolds $(M,H,{\cal J})$. Only those with   
pointwise invertible $P$ are. For these, $H=0$ and $\cal J$ is of the form (2.22) 
with $Q=-P^{-1}$. For this reason, the Hitchin sigma model is only a 
partial generalization of the Poisson sigma model since it can reproduce the 
latter only in the case where the target space Poisson bivector is pointwise 
invertible \ref{31}. 

Similarly, when $(M,P)$ is a Poisson manifold with pointwise invertible $P$, 
not all coisotropic submanifolds $W$ of $M$ are generalized complex submanifolds 
$(W,F)$ for some $F$ of the corresponding generalized complex manifold
$(M,H=0,{\cal J})$ with $\cal J$ of the form indicateded above. 
If we require, as it is reasonable to do, that $F=0$, then $W$ must be a Lagrangian 
submanifold of $M$ in order to be generalized complex submanifold, a more 
restrictive condition than being a coisotropic submanifold. 
This follows easily from (3.9{\it a, c}), by inspection. So, again, the brane Hitchin 
sigma model is only a partial generalization of the brane Poisson sigma model. 

As to the boundary conditions, the conditions (5.17{\it a, b}) with $F_{ij}=0$
reproduce those of ref. \ref{39} as already noticed. We have, however, the 
extra condition (5.25) with $J^i{}_j=0$ and $F_{ij}=0$. This condition follows from 
the requirement that the boundary integral (5.23) vanishes. Of course, when $F=0$, 
this condition is no longer strictly necessary.

\vskip.6cm\par\noindent{\bf Acknowledgments.}  
We thank F. Bastianelli for support and encouragement 
and M. Gualtieri for correspondence.

\vfill\eject   
\vskip.6cm
\vskip.6cm
\centerline{\bf REFERENCES}
\vskip.3cm

\item{[1]}
C.~Vafa,
``Superstrings and topological strings at large N'',
J.\ Math.\ Phys.\  {\bf 42} (2001) 2798,
\item{}
[arXiv:hep-th/0008142].

\item{[2]}
S.~Kachru, M.~B.~Schulz, P.~K.~Tripathy and S.~P.~Trivedi,
``New supersymmetric string compactifications'',
JHEP {\bf 0303} (2003) 061,
\item{} 
[arXiv:hep-th/0211182].

\item{[3]}
S.~Gurrieri, J.~Louis, A.~Micu and D.~Waldram,
``Mirror symmetry in generalized Calabi-Yau compactifications'',
Nucl.\ Phys.\ B {\bf 654} (2003) 61, 
\item{}
[arXiv:hep-th/0211102].

\item{[4]}
S.~Fidanza, R.~Minasian and A.~Tomasiello,
``Mirror symmetric SU(3)-structure manifolds with NS fluxes'',
\item{}
arXiv:hep-th/0311122.

\item{[5]}
M.~Grana, R.~Minasian, M.~Petrini and A.~Tomasiello,
``Supersymmetric backgrounds from generalized Calabi-Yau manifolds'',
JHEP {\bf 0408} (2004) 046,
\item{}
[arXiv:hep-th/0406137].

\item{[6]}
N.~Hitchin, 
``Generalized Calabi Yau manifolds'',
Q. \ J. \ Math. {\bf 54} no. 3 (2003) 281, 
\item{}
[arXiv:math.dg/0209099].

\item{[7]}
M.~Gualtieri,
``Generalized complex geometry'',
Oxford University doctoral thesis,
\item{}
arXiv:math.dg/0401221.

\item{[8]}
E.~Witten,
``Topological Sigma Models'',
Commun.\ Math.\ Phys.\  {\bf 118} (1988) 411.

\item{[9]}
E.~Witten,
``Mirror manifolds and topological field theory'',
in ``Essays on mirror manifolds'', ed. S.~T. ~Yau, International
Press, Hong Kong, (1992) 120,
\item{}
[arXiv:hep-th/9112056].

\item{[10]}
A.~Kapustin,
``Topological strings on noncommutative manifolds'',
IJGMMP {\bf 1} nos. 1 \& 2 (2004) 49,
\item{}
[arXiv:hep-th/0310057].

\item{[11]}
O.~Ben-Bassat,
``Mirror symmetry and generalized complex manifolds'',
\item{}
arXiv:math.ag/0405303.

\item{[12]}
C.~Jeschek,
``Generalized Calabi-Yau structures and mirror symmetry'',
\item{}
arXiv:hep-th/0406046.

\item{[13]}
A.~Kapustin and Y.~Li,
``Topological sigma-models with H-flux and twisted generalized complex manifolds'',
\item{}
arXiv:hep-th/0407249.

\item{[14]}
S.~Chiantese, F.~Gmeiner and C.~Jeschek,
``Mirror symmetry for topological sigma models with generalized Kahler geometry'',
\item{}
arXiv:hep-th/0408169.

\item{[15]}
S.~Chiantese,
``Isotropic A-branes and the stability condition'',
\item{}
arXiv:hep-th/0412181.

\item{[16]}
M.~Kontsevich, 
``Homological Algebra of Mirror Symmetry'', 
\item{}
arXiv:alg-geom/9411018.

\item{[17]}
M.~Douglas, 
``D-branes, Categories and N=1 Supersymmetry'', 
J.\ Math.\ Phys. {\bf 42} (2001) 2818,
\item{}
[arXiv:hep-th/0011017].

\item{[18]}
P.~P.~Aspinwall and A.~ Lawrence, 
``Derived Categories and Zero-Brane Stability'', 
JHEP {\bf 0108} (2001) 004,
\item{} 
[arXiv:hep-th/0104147].

\item{[19]}
E.~Witten, 
``Chern-Simons Gauge Theory as a String Theory'', 
Prog.\ Math. {\bf 133} (1995) 637,
\item{}
[arXiv:hep-th/9207094].

\item{[20]}
K.~Hori
``Mirror symmetry'',
Clay Mathematics Monographs {\bf 1}, American Mathematical Society, (2003).

\item{[21]}
H. Ooguri, Y. Oz and Z. Yin, 
``D-branes on Calabi-Yau Spaces and Their Mirrors'', 
Nucl.\ Phys.\ B {\bf 477} (1996) 407,
\item{}
[arXiv:hep-th/9606112].

\item{[22]}
A.~Kapustin and D.~Orlov,
J.\ Geom.\ Phys.\  {\bf 48} (2003) 84,
\item{}
[arXiv:hep-th/0109098].

\item{[23]}
M.~Zabzine,
``Geometry of D-branes for general N=(2,2) sigma models'', 
\item{}
arXiv:hep-th/0405240.

\item{[24]}
U.~Lindstrom,
``Generalized N = (2,2) supersymmetric non-linear sigma models'',
Phys.\ Lett.\ B {\bf 587} (2004) 216,
\item{}
[arXiv:hep-th/0401100].

\item{[25]}
U.~Lindstrom, R.~Minasian, A.~Tomasiello and M.~Zabzine,
``Generalized complex manifolds and supersymmetry'',
\item{}
arXiv:hep-th/0405085.

\item{[26]}
U.~Lindstrom,
``Generalized complex geometry and supersymmetric non-linear sigma models'',
\item{}
arXiv:hep-th/0409250.

\item{[27]}
U.~Lindstrom, M.~Rocek, R.~von Unge and M.~Zabzine,
``Generalized Kaehler geometry and manifest N = (2,2) supersymmetric nonlinear
sigma-models'',
\item{}
arXiv:hep-th/0411186.

\item{[28]}
A.~Alekseev and T.~Strobl,
``Current algebra and differential geometry'',
\item{}
arXiv:hep-th/0410183.

\item{[29]}
A.~Kotov, P.~Schaller and T.~Strobl,
``Dirac sigma models'',
\item{}
arXiv:hep-th/0411112.

\item{[30]}
L.~Bergamin,
``Generalized complex geometry and the Poisson sigma model'',
\item{}
arXiv:hep-th/0409283.

\item{[31]}
R.~Zucchini,
``A sigma model field theoretic realization of Hitchin's generalized complex
geometry'', JHEP {\bf 0411} (2004) 045,
\item{}
[arXiv:hep-th/0409181].

\item{[32]}
N.~Ikeda,
``Three Dimensional Topological Field Theory induced from Generalized Complex
Structure'',
\item{}
arXiv:hep-th/0412140.

\item{[33]}
N.~Ikeda,
``Two-dimensional gravity and nonlinear gauge theory'',
Annals Phys.\  {\bf 235} (1994) 435, 
\item{}
[arXiv:hep-th/9312059].

\item{[34]}
P.~Schaller and T.~Strobl,
``Poisson structure induced (topological) field theories'',
Mod.\ Phys.\ Lett.\ A {\bf 9} (1994) 3129, 
\item{}
[arXiv:hep-th/9405110].

\item{[35]}
A.~S.~Cattaneo and G.~Felder,
``A path integral approach to the Kontsevich quantization formula'',
Commun.\ Math.\ Phys.\  {\bf 212} (2000) 591, 
\item{}
[arXiv:math.qa/9902090].

\item{[36]}
I.~A.~Batalin and G.~A.~Vilkovisky,
``Gauge Algebra And Quantization'',
Phys.\ Lett.\ B {\bf 102} (1981) 27.

\item{[37]}
I.~A.~Batalin and G.~A.~Vilkovisky,
``Quantization Of Gauge Theories With Linearly Dependent Generators'',
Phys.\ Rev.\ D {\bf 28} (1983) 2567
(Erratum-ibid.\ D {\bf 30} (1984) 508).

\item{[38]}
A.~S.~Cattaneo and G.~Felder,
``On the AKSZ formulation of the Poisson sigma model'',
Lett.\ Math.\ Phys.\  {\bf 56} (2001) 163
\item{}
[arXiv:math.qa/0102108].

\item{[39]}
A.~S.~Cattaneo and G.~Felder,
``Coisotropic submanifolds in Poisson geometry and branes in the Poisson sigma model'',
Lett.\ Math.\ Phys.\  {\bf 69} (2004) 157
\item{}
[arXiv:math.qa/0309180].

\item{[40]}
T. Courant,
``Dirac manifolds'',
Trans. \ Amer. \ Math. \ Soc. {\bf 319} no. 2 (1990) 631.
  
\item{[41]}
T. Courant and A. Weinstein,
``Beyond Poisson structures'',
in ``Action hamiltoniennes des groupes, troisi\`eme th\'eor\`eme de Lie'',
Lyon (1986) 39, Travaux en Cours 27, Hermann, Paris (1988). 

\item{[42]}
J.~-L.~Brylinski,
``Loop Spaces, Characteristic Classes and Geometric Quantization'',
Birkh\"auser, 1993.

\item{[43]}
S.~J.~Gates, C.~M.~Hull and M.~Rocek,
``Twisted Multiplets And New Supersymmetric Nonlinear Sigma Models'',
Nucl.\ Phys.\ B {\bf 248}, 157 (1984).

\item{[44]}
O.~Ben-Bassat and M.~Boyarchenko,
``Submanifolds of generalized complex manifolds'',
\item{}
arXiv:math.dg/0309013.

\item{[45]}
L.~Baulieu, A.~S.~Losev and N.~A.~Nekrasov,
``Target space symmetries in topological theories I'',
JHEP {\bf 0202} (2002) 021,
\item{}
[arXiv:hep-th/0106042].

\item{[46]}
R.~Zucchini,
``Target space equivariant cohomological structure of the Poisson sigma model'',
J.\ Geom.\ Phys.\  {\bf 48} (2003) 219,
\item{}
[arXiv:math-ph/0205006].

\item{[47]}
P.~A.~Grassi and A.~Quadri,
``The background field method and the linearization problem for Poisson manifolds'',
\item{}
arXiv:hep-th/0403265.

\item{[48]}
C.~Klimcik and T.~Strobl,
``WZW-Poisson manifolds'',
J.\ Geom.\ Phys.\  {\bf 43} (2002) 341,
\item{}
[arXiv:math.sg/0104189].

\item{[49]}
J.~Cheeger,
``Multiplication of Differential Characters'',
Convegno Geometrico INDAM, Roma maggio 1972, in 
Symposia Mathematica {\bf XI} Academic Press (1973) 441.

\item{[50]}
J.~Cheeger and J.~Simons,
``Differential Characters and Geometric Invariants'',
Stony Brook preprint (1973) reprinted in 
Lecture Notes in Math. {\bf 1167} Sprin\-ger Verlag (1985) 50.

\item{[51]}
R.~Zucchini,
``Relative topological integrals and relative Cheeger-Simons differential
characters'',
J.\ Geom.\ Phys.\  {\bf 46} (2003) 355
\item{}
[arXiv:hep-th/0010110].

\item{[52]}
M.~Alexandrov, M.~Kontsevich, A.~Schwartz and O.~Zaboronsky,
``The Geometry of the master equation and topological quantum field theory'',
Int.\ J.\ Mod.\ Phys.\ A {\bf 12} (1997) 1405
\item{}  \
[arXiv:hep-th/9502010].

\vfill\eject

\item{[53]}
I.~Calvo and F.~Falceto,
``Poisson reduction and branes in Poisson-sigma models'',
\item{}
arXiv:hep-th/0405176.

\item{[54]}
I.~Calvo and F.~Falceto,
``Poisson-Dirac branes in Poisson-sigma models'',
\item{}
arXiv:hep-th/0502024.


\bye

The odd vector field is the sum of the commuting vector fields $\hat D$ and 
$\check Q$ obtained from $D$ and $Q$ by acting on maps 
$\Pi T\Sigma\to\Pi T^*M$ by the corresponding
infinitesimal diffeomorphisms of $\Pi T\Sigma$ on the left and
of $\Pi T^*M$ on the right. The Hamiltonian function of this vector
field is then the BV action functional, if $\Sigma$
has no boundary. It is a function on
the space of maps from $\Pi T\Sigma$ to $\Pi T^*M$, whose Poisson
bracket with itself vanishes, i.e., it solves the BV
classical master equation. If $\Sigma$ has a boundary,
suitable boundary conditions must be 
imposed. They are first-class constraints, and the correct action
functional is obtained after Hamiltonian reduction.

In the case of the Poisson sigma model,
the construction goes as follows: to a
an oriented two-dimensional manifold $\Sigma$, possibly with boundary, one
associates the supermanifold $\Pi T\Sigma$, the tangent bundle
with reversed parity of the fiber. The algebra of functions on
$\Pi T\Sigma$ is isomorphic to 
the (graded commutative) algebra of differential forms 
on $\Sigma$ with values in a graded commutative ground ring $\Lambda$.
The integration of differential forms and the de~Rham differential
are, in the language of supermanifolds, a measure $\mu$ and a self-commuting
vector field $D$, respectively, on $\Pi T\Sigma$.

\bye